# Towards reproducible machine learning-based process monitoring and quality prediction research for additive manufacturing

**Jiarui Xie [a], Mutahar Safdar [a], Andrei Mircea [a], Bi Cheng Zhao [a], Yan Lu [b], Hyunwoong Ko [c], Zhuo Yang [b], Yaoyao Fiona Zhao [a,*]**

[a] Department of Mechanical Engineering, McGill University, QC, Canada
[b] Information Modeling and Testing Group, National Institute of Standards and Technology, MD, USA
[c] School of Manufacturing Systems and Networks, Arizona State University, AZ, USA
[*] Corresponding author.
Email address: yaoyao.zhao@mcgill.ca (Y. F. Zhao)

## ABSTRACT

Machine learning (ML)-based cyber-physical systems (CPSs) have been extensively developed to improve the print quality of additive manufacturing (AM). However, the reproducibility of these systems, as presented in published research, has not been thoroughly investigated due to a lack of formal evaluation methods. Reproducibility, a critical component of trustworthy artificial intelligence, is achieved when an independent team can replicate the findings or artifacts of a study using a different experimental setup and achieve comparable performance. In many publications, critical information necessary for reproduction is often missing, resulting in systems that fail to replicate the reported performance. This paper proposes a reproducibility investigation pipeline and a reproducibility checklist for ML-based process monitoring and quality prediction systems for AM. The pipeline guides researchers through the key steps required to reproduce a study, while the checklist systematically extracts reproducibility-relevant information from the publication. We validated the proposed approach through two case studies: reproducing a fused filament fabrication warping detection system and a laser powder bed fusion melt pool area prediction model. Both case studies confirmed that the pipeline and checklist successfully identified missing information, improved reproducibility, and enhanced the performance of reproduced systems. Based on the proposed checklist, a reproducibility survey was conducted to assess the current reproducibility status within this research domain. By addressing this research gap, the proposed methods aim to enhance trustworthiness and rigor in ML-based AM research, with potential applicability to other ML-based CPSs.

## 1. INTRODUCTION

Additive manufacturing (AM) has revolutionized the design and production sectors by facilitating rapid prototyping and enabling mass customization [1, 2]. Additionally, AM supports sustainability efforts by minimizing material waste and allowing for the recycling of components at the end of their lifecycle [3, 4]. However, despite its promising advantages, AM technologies are still evolving and often lack reliability [5]. The underlying physics of rapid phase transitions, such as grain growth, in AM processes remains insufficiently understood [6]. Moreover, the manufacturing process is prone to various anomalies that can negatively impact the quality of the final parts [5]. Consequently, even when identical designs and process parameters are employed, a batch of AM parts may exhibit varying levels of conformity.

To address the reliability challenges in AM, machine learning (ML)-based process monitoring and quality prediction in AM (PQ-AM) have been extensively researched [7]. Drawing inspiration from cyber-physical systems (CPSs), researchers collect data from AM systems during or after manufacturing processes using sensors like digital microphones and pyrometers [8]. Salient features can be extracted from the collected data and correlated with the associated labels such as defect types and part quality. For instance, Khan et al. [9] and Bisheh et al. [10] built fused filament fabrication (FFF) defect detection systems to detect infill pattern defects and geometric deviations using convolutional neural network (CNN) and gradient boosting, respectively. Snow et al. [11] and Jiang et al. [12] collected in-situ FFF images using cameras and trained deep neural networks to predict the bonding quality.

While PQ-AM systems hold significant promise for enhancing the reliability of AM processes, their reproducibility has yet to be thoroughly examined. Reproducibility in scientific research refers to the capacity for others to replicate a study and achieve the same results [13]. While it is widely recognized as a fundamental aspect of trustworthy artificial intelligence (AI) [14], ML-based methods have faced significant challenges regarding reproducibility across various engineering domains. A cross-disciplinary survey reported by Baker [15] indicated that 90% of researchers acknowledge a reproducibility crisis, with over 60% encountering difficulties when attempting to replicate published studies. In the manufacturing field, researchers primarily contribute to the community by publishing their work to demonstrate the feasibility and effectiveness of their proposed methodologies or systems. Publications serve as a critical avenue for academic researchers to build upon existing studies and for industrial practitioners to apply published methods to their projects. Therefore, publishing is the foremost means of disseminating information about original research and reproducibility is vital for claimed contributions [16]. Publications that include the necessary information for reproducibility enable readers to replicate the proposed methods, thereby enhancing the impact and credibility of the authors. Furthermore, surveys by Pineau et al. [17] have shown that papers with higher levels of reproducibility are more likely to be accepted for publication.

ML-based PQ-AM is inherently interdisciplinary, requiring expertise in both AM and ML, which adds to the complexity of ensuring reproducibility. In addition to providing comprehensive details about the AM and monitoring technologies used, it is essential to present information on the dataset and model [17]. Other crucial reproducibility details include, but are not limited to, experimental design, data preparation, model structure, and model selection [17-19]. Indeed, many ML-based monitoring studies in manufacturing, such as machine fault diagnostics and production line scheduling, encounter reproducibility challenges due to their interdisciplinary nature. The existing methods, such as the predictive model markup language (PMML) [20] and the FAIR principles [21], only partially and indirectly enhance the reproducibility of ML-based process or condition monitoring systems, without incorporating engineering domain knowledge. To the best of our knowledge, there is currently no established guideline or checklist for ensuring reproducibility in ML-based CPSs that integrate both ML and engineering domain knowledge.

This paper introduces a pipeline that systematically examines the reproducibility of ML-based in-process and in-situ PQ-AM (IIPQ-AM) systems, utilizing the CRoss Industry Standard Process (CRISP) methodology, which serves as a process model for data mining projects [22]. Additionally, we propose an ML-based IIPQ-AM system reproducibility checklist designed to assist authors in meeting reproducibility requirements. We investigated the reproducibility of two ML-based IIPQ-AM publications and reproduced their models using the proposed checklist and pipeline. Thereafter, we surveyed the level of reproducibility in this domain using the SciHIT IE (Scientific Human-AI Teaming for Information Extraction) platform developed by the Additive Design and Manufacturing Lab (ADML) and empowered by large language models (LLMs). The contributions of this paper are:

- We define the reproducibility, repeatability, and replicability of ML-based IIPQ-AM systems.

- We propose an ML-based IIPQ-AM reproducibility investigation pipeline based on the CRISP methodology for investigating the reproducibility of IIPQ-AM systems.
- We design an ML-based IIPQ-AM reproducibility checklist to help authors achieve high reproducibility for their original systems.
- We validate the reproducibility pipeline and the checklist by demonstrating their effectiveness in two case studies.
- We survey the reproducibility status of ML-based IIPQ-AM research using the proposed checklist and the SciHIT IE platform empowered by LLMs.

The remainder of this paper is organized as indicated in Figure 1. Section 2 provides a review of current research on reproducibility in both AM and ML, and subsequently defines the reproducibility of ML-based IIPQ-AM. Section 3 details the reproducibility checklist and outlines the reproducibility investigation pipeline. We validated the proposed method in section 4 by reproducing two ML-based IIPQ-AM research works, including an FFF warp detection model and a laser powder bed fusion (LPBF) melt pool area prediction model. Section 5 conducts a reproducibility survey using the checklist and the SciHIT IE platform. Section 6 highlights the remarks of this research.

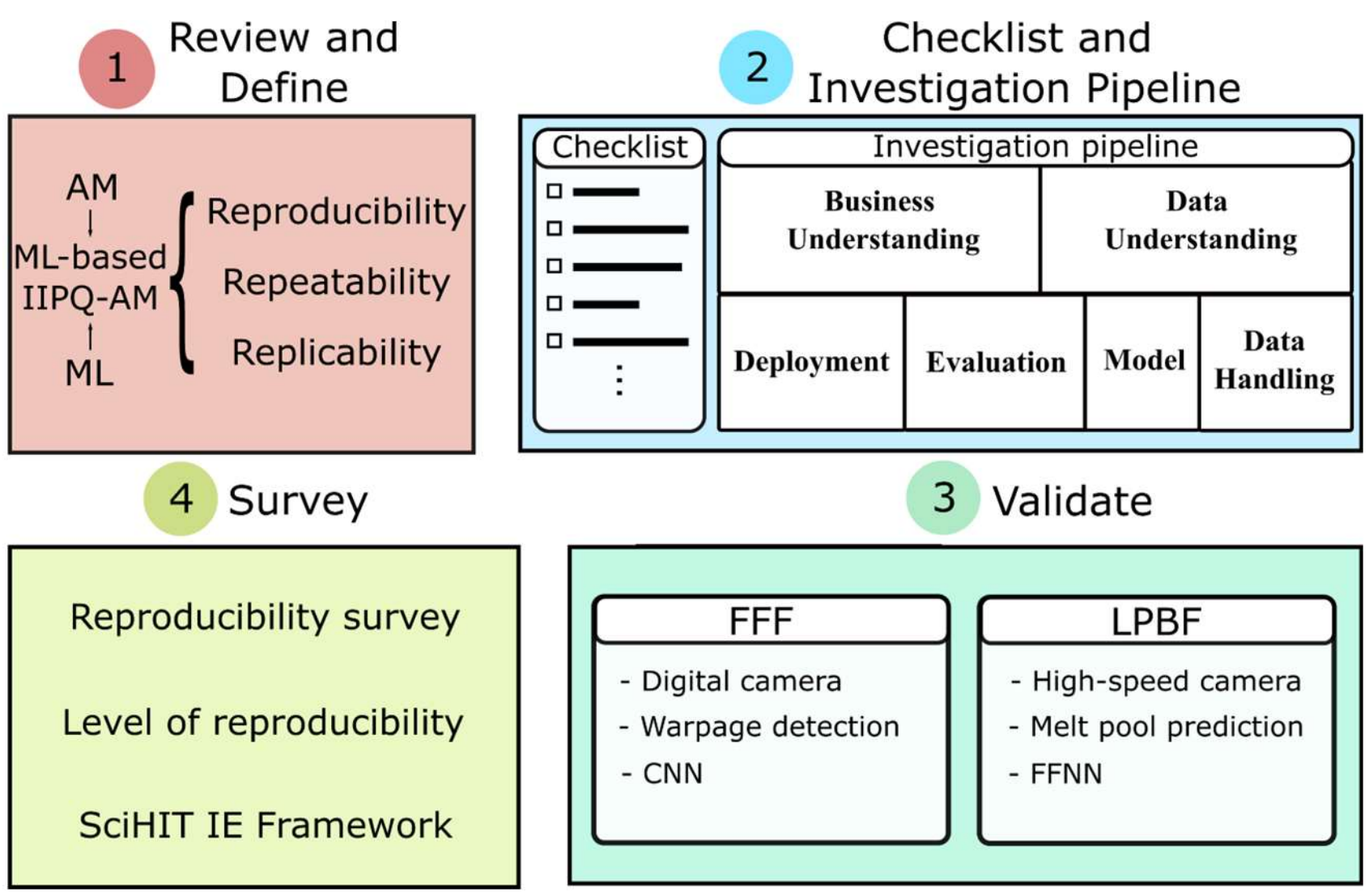

**Figure 1: Structure of the manuscript.**

## 2. BACKGROUND

This section reviews the definitions of reproducibility, repeatability, and replicability in related domains and provides the proposed definitions for ML-based IIPQ-AM systems. By surveying relevant frameworks and techniques, we found that there has not been a comprehensive and suitable method to investigate and enhance the reproducibility of ML-based IIPQ-AM systems, thus justifying the necessity of the proposed investigation pipeline and checklist.

### 2.1. Reproducibility, repeatability, and replicability in relevant domains

Reproducibility is a crucial measure of the transparency and authenticity of a published work. Although reproducibility definitions and rules have been established, the lack of reproducibility in scientific research is becoming a serious challenge for scientific discovery. Baker [15] distributed a reproducibility survey to multiple domains such as chemistry and physics and obtained responses from 1576 researchers. It revealed that more than 60% of the respondents experienced failure when reproducing others' results, although 66% claimed that reproducibility protocols have been established in their laboratories. A lack of reproducibility was also revealed by surveys in biomedical [23], cancer biology [24], and psychology [25] domains.

Reproducibility, repeatability, and replicability have been defined and investigated for AM and ML. Although these terms all aim to achieve the same results or replicate the original methods, they apply under different conditions (Table 1). In AM, repeatability refers to achieving the same precision when manufacturing an identical part by the same operator using the same equipment, whereas reproducibility involves different operators and equipment [26]. Pineau et al. [17] defined ML reproducibility and replicability as producing the same results using the same model with the same and different data, respectively. ML-based IIPQ-AM systems fall under the category of artifacts, which are physical or digital objects created by human beings [27]. Association for Computing Machinery (ACM) [27] differentiates the artifact reproducibility, repeatability, and replicability according to the group and experiment setup. Drawing from the AM and ML domains, the definitions of these three terminologies have been adapted for ML-based IIPQ-AM and are presented in Table 1.

**Table 1: Definitions of reproducibility, repeatability, and replicability in relevant domains.**

| Domain | Reproducibility | Repeatability | Replicability |
|---|---|---|---|
| **AM [26]** | Same precision<br>Different operator<br>Different equipment | Same precision<br>Same operator<br>Same equipment | Not defined |
| **ML [17]** | Same result<br>Same data<br>Same model | Not defined | Same result<br>Different data<br>Same model |
| **Artifact [27]** | Same result<br>Different group<br>Different experiment setup | Same result<br>Same group<br>Same experiment setup | Same result<br>Different group<br>Same experiment setup |
| **ML-based IIPQ-AM [This work]** | Comparable performance<br>Different group<br>Different experiment setup | Comparable performance<br>Same group<br>Same experiment setup | Comparable performance<br>Different group<br>Same experiment setup |

In AM, reproducibility typically refers to the ability to consistently manufacture a part with specific desired characteristics, such as its structure and properties. Kim et al. [28] defined producibility, repeatability, and reproducibility in AM in relation to the information elements within the AM digital thread. Producibility pertains to the fundamental information required to produce a part. Repeatability includes process-related information elements, both system-dependent and system-independent, and focuses on the ability to repeat the same AM process without considering the resulting characteristics. Part reproducibility expands the information thread to encompass elements of structure and/or property, ensuring that the manufactured part meets these specifications. The concepts of repeatability and reproducibility also apply to AM techniques for quality inspection and part evaluation, which generate data for ML-based IIPQ-AM. In a related work, Seo et al. [29] referred to repeatability as process repeatability and reproducibility as part reproducibility to present a quality assurance framework in AM. Cacace et al. [30] evaluated the repeatability (e.g., consecutive measurements) and reproducibility (e.g., measurements after system disassembly) for an X-ray Computed Tomography system. Gunay et al. [31] defined reproducibility in AM as the ability to produce parts under the same conditions. In their work, reproducibility was closely related to dimensional accuracy, but this could change depending on the AM application. Similarly, Petrovic et al.

[32] highlighted the challenges of achieving part reproducibility in AM, attributing it to inadequate quality and testing standards.

ML reproducibility faces unique challenges from both the data and the model involved. Raff [33] independently implemented the ML algorithms proposed in 255 manuscripts, in which 63.5% of the manuscripts were considered reproducible. However, this study adopted several reproducibility criteria that are less stringent than the ones proposed in other reproducibility studies [23-25]. To help improve the reproducibility of ML research, Pineau et al. [17] proposed an ML reproducibility checklist that highlights critical reproducibility information with respect to the model, dataset, code, experimental results, and theoretical claim of a paper. Filling up the reproducibility checklist has been incorporated as a submission requirement for the Conference on Neural Information Processing Systems (NeurIPS) since 2018. Moreover, Reproducibility Challenges have been held since 2018 in prestigious ML conferences, including the International Conference on Learning Representations (ICLR) and NeurIPS [17, 34]. Attendees of these events attempt to reproduce the results of the papers accepted by the conferences and report their attempts. Reproducibility Challenges have attracted thousands of participants and raised awareness of ML reproducibility.

### 2.2. Reproducibility of ML-based IIPQ-AM systems

This paper focuses on the reproducibility of ML-based IIPQ-AM systems, a joint domain of AM and ML. An ML-based IIPQ-AM system typically consists of four key elements:

1) **Manufacturing system**: The manufacturing system encompasses the manufacturing process technology, the associated hardware, and the printing material(s). The hardware and materials may either be standard, commercially available components or custom-designed specifically for a given system.
2) **Sensing system**: The sensing system represents the sensing or monitoring technology, associated hardware, and configuration.
3) **Dataset**: Data gathered from the manufacturing and sensing systems, together with the methods used to process and prepare this data for ML.
4) **Model**: The proposed ML model, including its hyperparameters, training methods, and model selection techniques.

Adapted from the existing reproducibility definitions in AM and ML, the definition of reproducibility for ML-based IIPQ-AM systems is proposed: To fulfill the reproducibility requirements, an ML-based IIPQ-AM system should be reproduced (Table 1):

- **With a different experiment setup**: The experimental setup used in the original manufacturing and sensing systems may not be ideal or readily accessible for reproduction purposes. A clear description of the experiment setup allows the reader to understand and develop their own system.
- **By a different group**: The reproduction procedure is carried out by a different group that was not involved in the original development of the system. This new group may face challenges due to a lack of information or expertise when attempting to reproduce the system.
- **To obtain comparable performance**: Reproducing ML-based IIPQ-AM systems involves challenges related to both hardware and software components. Rather than striving to replicate the reported performance exactly, it is more practical to achieve performance that is comparable to that of the original systems.

Despite the rapid growth in ML-based IIPQ-AM publications, only a limited number of works indirectly investigate and enhance the reproducibility in this domain. Ko et al. [35] proposed the Data-Knowledge-Design Rule (DKDR) framework that integrates ML and knowledge graphs to formalize unstructured AM guidelines into structured knowledge, design rules, and ontologies. It transforms AM data from the National Institute of Standards and Technology (NIST) AM Materials Database (AMMD) [36] into actionable knowledge and develops predictive models for AM design and manufacturability. The framework enhances the reproducibility of ML-based AM quality prediction models by ensuring standardized data use and clear design-rule relationships. However, DKDR focuses solely on data-driven design, without specifying the configurations of the sensing system and the ML models. Safdar et al. [37] proposed a transferability analysis framework for data-driven AM knowledge consisting of six AM knowledge components (AM process, material, system, model, activity, and concern) and five ML knowledge components (ML task, model, input, preprocessing, and output). This framework can help analyze the reproducibility of ML-based IIPQ-AM systems according to the AM and ML knowledge components. However, the knowledge components are not designed to cover the details of the four key elements, especially for the sensing system and dataset. The FAIR principles (Findable, Accessible, Interoperable, Reusable) are a set of guidelines designed to enhance the handling and sharing of digital assets, ensuring they are simple to locate, obtain, integrate, and reuse across various systems [21]. Incorporating the FAIR principles into the development and sharing of ML-based AM monitoring systems will lead to more transparent and reproducible research, with well-defined and accessible datasets, models, and workflows. The PMML is an extensible markup language (XML)-based standard that allows for the sharing and deployment of trained ML models across different platforms [20]. With the model structure, algorithms, and parameters transparently documented, the PMML allows for the exchange of models across various systems, ensuring that the same models can be replicated consistently. Nonetheless, the FAIR principles and PMML focus on the datasets and models, without embedding AM domain knowledge to incorporate the manufacturing and sensing systems. Besides, they do not capture some essential information for reproducing ML models, such as model training and hyperparameter tuning. In summary, the abovementioned frameworks and methods are not specifically designed to investigate the reproducibility of ML-based IIPQ-AM systems, thus falling short of covering all reproducibility considerations in this domain.

Without awareness and evaluation methods of reproducibility, this domain faces a rapidly increasing number of publications with unknown reproducibility status, leading to an urgent need for guidelines that allow readers and authors to verify and enhance the reproducibility of research. The subsequent sections will analyze ML-based IIPQ-AM systems, introduce a reproducibility investigation pipeline along with a checklist, assess the proposed method through two case studies, and conduct a reproducibility survey.

## 3. METHODOLOGY

This section examines ML-based IIPQ-AM systems and develops a reproducibility investigation pipeline and checklist. An ML-based IIPQ-AM reproducibility investigation pipeline is established based on the CRISP methodology. Following this, an ML-based IIPQ-AM reproducibility checklist is formulated to assess the reproducibility of each phase within the pipeline. The purpose of the proposed pipeline and checklist is to determine whether the manuscript provides all necessary information for reproducing the system. Given the specific objectives, this method must encompass all essential reproducibility requirements while avoiding excessive detail that could overwhelm users. Therefore, the approach must strike a balance between comprehensiveness and concision.

### 3.1. ML-based IIPQ-AM Reproducibility Investigation Pipeline

The CRISP methodology is employed to analyze ML-based IIPQ-AM systems and develop a reproducibility investigation pipeline for ML-based IIPQ-AM publications. As a well-established data mining process model, the CRISP methodology is applicable across various business contexts (Figure 2) [22].

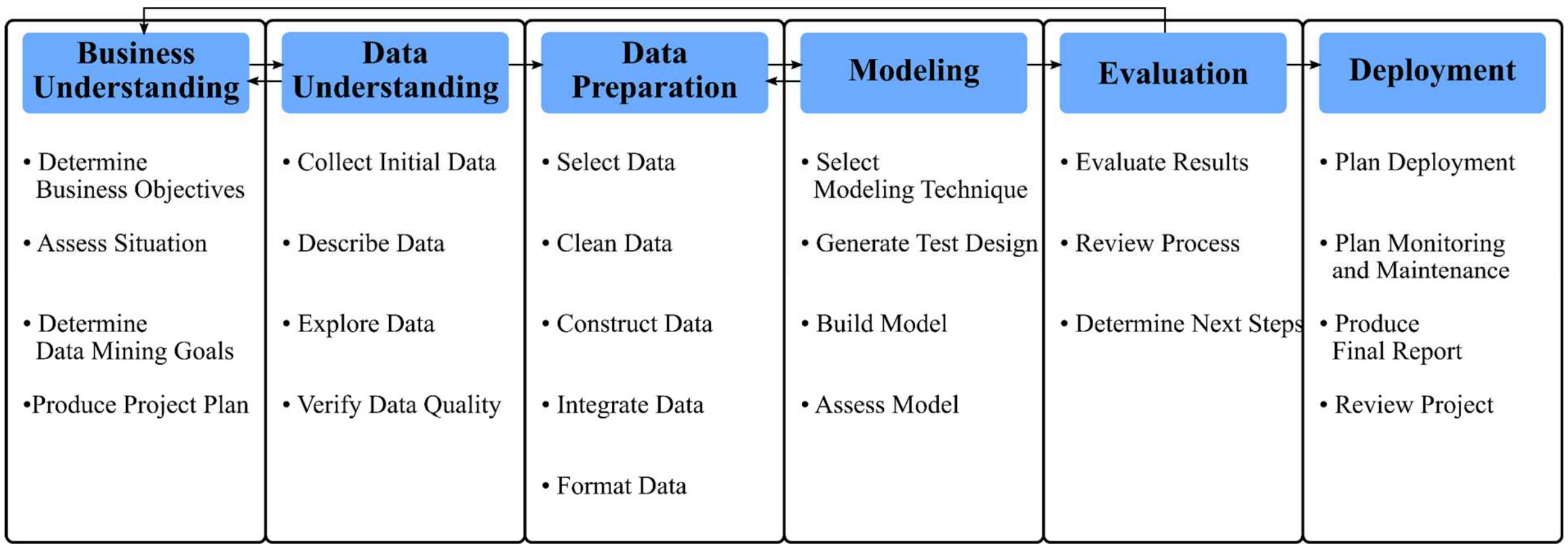

**Figure 2: Phases of the CRISP process model (adapted from [22]).**

The proposed reproducibility investigation pipeline divides the reproduction process into six phases based on the CRISP methodology (Figure 3). The objective of this pipeline is to assess the reproducibility of an IIPQ-AM system and facilitate the reproduction. In Figure 3, blue blocks represent IIPQ-AM hardware or software, white blocks represent hardware sub-elements, parallelograms indicate inputs or outputs, pink blocks represent processes, and blue arrows represent the direction of information flow. The following paragraphs elaborate on each phase of the CRISP methodology and the IIPQ-AM reproducibility investigation pipeline.

**Business Understanding.** The business understanding phase examines the project objectives and requirements from a business perspective. The outcomes of this phase provide the foundation for designing a data mining project and creating an initial project plan. In the context of an ML-based IIPQ-AM system, business understanding involves analyzing the manufacturing system and the modeling purpose (Figure 3a). This includes detailing the manufacturing system, such as the AM machine and materials, to highlight the key hardware components. The modeling purpose such as process anomaly detection and control must be discussed to illustrate the capability of the system. Besides, the physical phenomena being monitored must be discussed to bridge the manufacturing and sensing systems.

**Data Understanding.** The data understanding phase involves initial data collection and analysis to extract preliminary statistical insights and identify potential data quality issues. In the proposed investigation pipeline, this phase focuses on the sensing system and the raw datasets (Figure 3b). An ML-based IIPQ-AM system utilizes specific sensors based on the physical phenomena being monitored. Additionally, sensor calibrations and various supporting apparatuses can help filter out unwanted signals or highlight target signals to improve data quality. Data understanding also covers the experimental design that produces the raw dataset. Finally, the characteristics of the raw datasets collected through the sensing system, including data formats, modalities, and statistics, need to be thoroughly described.

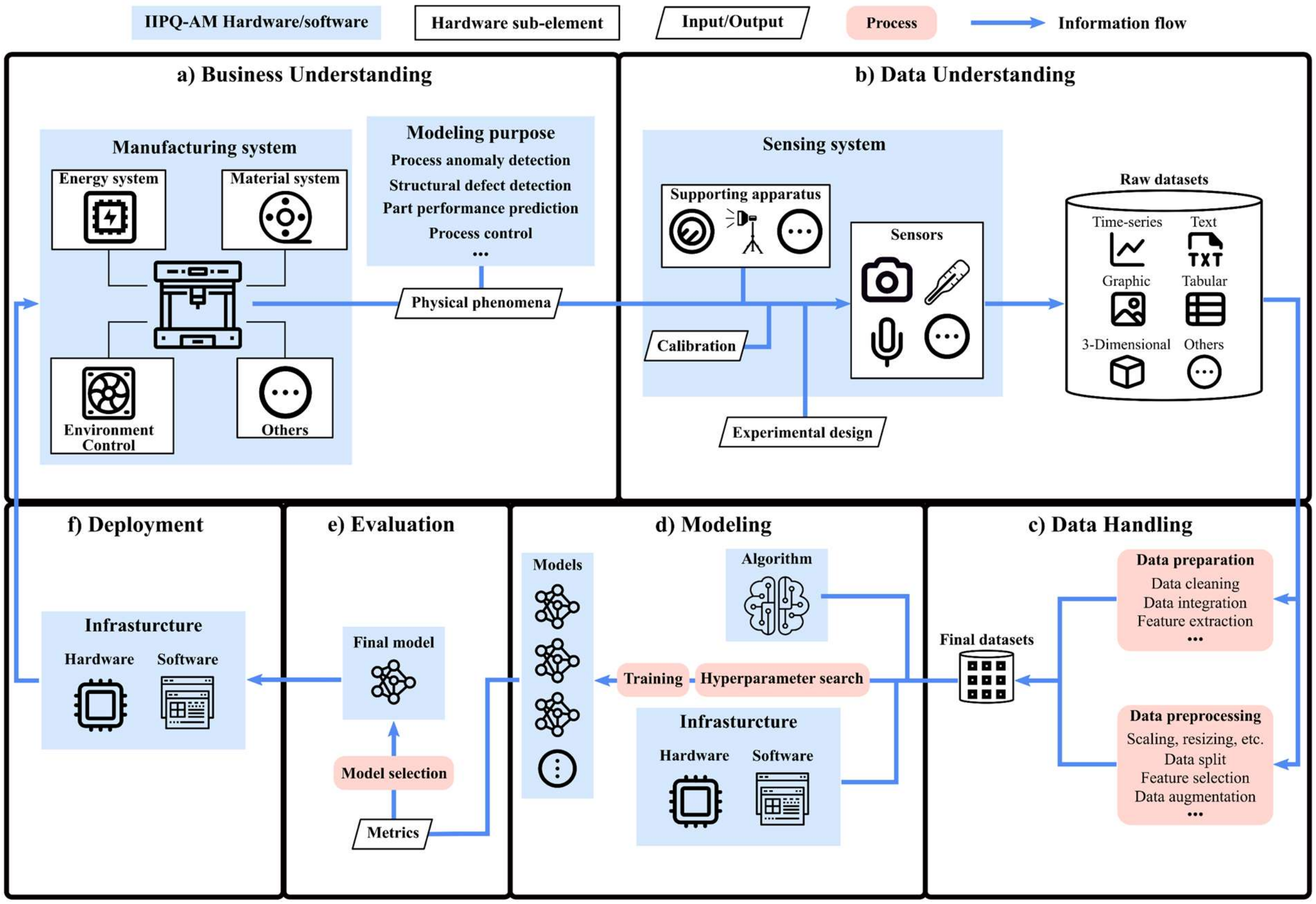

**Figure 3: ML-based IIPQ-AM reproducibility investigation pipeline based on the CRISP methodology.**

**Data Handling.** The data handling phase encompasses various data preparation and preprocessing techniques (Figure 3c). Data preparation techniques leverage engineering knowledge to transform and align raw datasets. Data preprocessing techniques transform the data into a suitable form for ML to improve the learning performance [19]. The final datasets used for training the ML models are produced following these transformations.

**Modeling.** The modeling phase involves selecting the modeling techniques, training the ML models, and preliminarily testing the models (Figure 3d). In the context of ML-based IIPQ-AM, this phase begins with selecting appropriate ML algorithms based on the specific modeling task and data modality. A hyperparameter search technique is then employed to find the optimal hyperparameters for each selected ML algorithm. Following this, a set of ML models is trained using the chosen algorithms and hyperparameters, utilizing the designated infrastructure.

**Evaluation.** The evaluation phase involves assessing the trained models and selecting the final model (Figure 3e). In this process, appropriate metrics must be established to evaluate the ML models based on the specific modeling task. A model selection strategy is then used to identify the best-performing model according to these metrics. Ultimately, the optimal performance of the ML-based IIPQ-AM system is determined during this evaluation phase.

**Deployment.** In this phase, the ML model is being deployed to make decisions and guide future AM activities (Figure 3f). A clear description of the hardware and software requirements is crucial for deploying the ML-based IIPQ-AM system. Failure to meet the requirements would compromise the model performance and/or cause interoperability issues.

### 3.2. ML-based IIPQ-AM Reproducibility Checklist

A deep dive analysis into each phase of the investigation pipeline results in the ML-based IIPQ-AM reproducibility checklist, composed of a series of reproducibility questions. Each reproducibility question is elaborated with definitions, discussions, and examples with references.

#### 3.2.1. Business Understanding

**Special notes on the AM system:** Many ML-based IIPQ-AM studies are expected to utilize well-established AM technologies, such as the seven standardized AM process techniques defined by the American Society for Testing and Materials (ASTM) [38]. An AM manufacturing system includes the specific AM technology in use and the associated components of its printing setup, including the materials used for printing parts. Gibson et al. [39] outlined major AM process technologies and their manufacturers, providing essential details for setting up a system to reproduce parts using a particular AM technology. In some cases, a customized manufacturing system requires additional reproducibility information. An AM system can include various subsystems, such as material feeding, energy supply, part handling or positioning, and environment control. The material feeding system facilitates the layer-by-layer addition of feedstock according to the process plan. All AM technologies rely on energy (either direct or indirect) to perform the build process. The handling system supports the substrate or under-build part during deposition, while the environment control system maintains specific conditions in the build environment for certain applications. Customization-specific system-level information is necessary for reproducing parts across different laboratories or production facilities, as highlighted in the checklist.

**RQ1. Is the manufacturing system encompassing process technology and associated hardware component(s) reported?**

**Explanation:** Given that most systems are anticipated to be off-the-shelf AM machines using standard process technologies from suppliers like Optomac, EOS, Renishaw, Desktop Metal, 3D Systems, and similar companies, the first question in the checklist focuses on describing the base manufacturing or printing system. Providing details about the base manufacturing system helps researchers identify the key manufacturing assets and their suppliers, which is crucial for reproducing the system.

**RQ2. If applicable, is customization of the manufacturing system reported?**

**Explanation:** In certain research and development scenarios, the printing setup may be customized by integrating specific components into the base manufacturing system. The second question in the reproducibility checklist addresses customized systems where components might be sourced from various suppliers or developed internally. It is essential to provide detailed information about these components—such as those related to energy, material feeding, part handling, cooling, and environmental control. Additionally, the scope of customization may include the materials used and the process technology employed.

**RQ3. Are characteristics of the material system reported?**

**Explanation:** The reproducibility checklist includes a question about the characteristics of the material system used to fabricate the test specimen. Authors should specify the material name, grade, and relevant specifications.

**RQ4. Is the modeling purpose discussed?**

**Explanation:** This question identifies the modeling task of the ML-based IIPQ-AM system. Common modeling tasks encompass AM process anomaly detection, structural defect identification, part performance prediction, and process control. Explaining the modeling purpose aids in justifying the choices of sensing system configurations, data handling methods, and ML models.

**RQ5. Are the physical phenomena being captured by the sensing system introduced?**

**Explanation:** This question connects the manufacturing and sensing systems. In ML-based IIPQ-AM, the input data captured by the sensing system are fed into the ML model to perform the prediction task. The sensing system captures physical phenomena occurring in the AM process, which provide crucial information related to the modeling task. Physical phenomena include both the AM process concern (e.g., melt pool or thermal gradient) and the signal (acoustic emission, vibration, or infrared light) being captured.

### 3.2.2. Data Understanding

**Special notes on the sensing system:** In ML-driven AM research, sensors are distinguished from base AM systems due to their crucial role in generating data for these applications. AM involves a complex range of multi-phase and multi-physics processes, resulting in a diverse array of sensors. Zhu et al. [8] reviewed condition monitoring in metal AM and categorized AM sensing technologies based on the types of signals they capture, including optical, thermal, acoustic, and vibration sensors, with optical and acoustic sensors being the most commonly used. In addition, Chen et al. [40] identified laser-line scanning and multi-sensor fusion as two prominent in-situ monitoring techniques in their recent survey of laser additive manufacturing (LAM). As highlighted by the data understanding phase (Figure 3b), detailed information about the sensing system is essential for reproducing ML-based IIPQ-AM systems. The reproducibility checklist covers sensor type and specifications, sensor settings and calibration, and sensor deployment.

**RQ6. Are sensor type(s) and specification(s) provided?**

**Explanation:** Sensor specification details are crucial for determining the appropriate sensors to use when reproducing the system [41]. Using sensors with similar capacity and capability helps ensure data are collected with comparable fidelity, which increases the likelihood of successful reproduction. Common sensor types in AM include optical cameras, infrared cameras, photodiodes, pyrometers, acoustic emission sensors, microphones, ultrasonic transducers, and accelerometers [8]. If the original system uses off-the-shelf sensors, their serial numbers can often provide the necessary specifications. If serial numbers are not available, a detailed set of specifications should be provided to ensure similar data fidelity. For instance, to describe a high-speed camera, specifications such as pixel pitch, image size, frame rate, and shutter speed ranges should be included.

**RQ7. Are the sensor settings or calibration details discussed?**

**Explanation:** Details on how sensors are adjusted and calibrated for specific sensing tasks must be included [42]. Sensors can be fine-tuned or calibrated to capture the relevant signals effectively. For example, the sampling rate of a high-speed camera should be calibrated to match the manufacturing process; rates that are too high or too low can lead to excessive or insufficient data. The camera resolution should be set according to the region of interest (ROI), with resolutions that are either too large or too small potentially causing noise or missing information. Other typical settings and calibrations include zoom, brightness, sampling rate, contrast ratio, and shutter speed.

**RQ8. Are the sensor deployment details discussed?**

**Explanation:** The physical configuration of the sensing system must be thoroughly detailed, including the attachment of sensors to the manufacturing system and the arrangement of any supporting equipment [43]. For instance, information about the location, angle, and field of view of a camera should be provided. Additionally, the placement, angle, and intensity of any light sources must be discussed. If filters are used, their positions and functions should also be explained.

**RQ9. Are the data modalities, formats, and types introduced?**

**Explanation:** The modalities, formats, and types of both raw and final datasets should be described [7]. For instance, datasets can be in various forms such as tabular, graphical, 3-dimensional, time-series, or textual. Tabular datasets might come in formats like .csv, .txt, or .xlsx. Additionally, the types of variables within the datasets should be specified, such as nominal, ordinal, discrete, or continuous. Whenever the data is represented in a transformed format, such as false color maps, the scale used should be provided. This ensures that the readers can accurately interpret the transformed data and understand the relationship between the original data and its transformed representation. Providing clear details on data formats and types enhances compatibility when reproducing the systems.

**RQ10. Are the relevant statistics provided?**

**Explanation:** The statistics of the raw and the final datasets such as data quantity, number of input variables, and number of classes should be discussed if applicable. Since data form the foundation of ML-based modeling, detailed statistics should be provided for the raw data, data post-transformations, and the final dataset. This information is crucial for understanding how datasets evolve through transformations and for identifying potential issues such as data imbalance or scarcity that could impact the training and evaluation of ML models.

**RQ11. Are the experimental design settings for data acquisition introduced?**

**Explanation:** The values of the experimental parameters used in designing experiments for data acquisition must be detailed, including both fixed and variable parameters. These parameters can be categorized into part design parameters and AM process parameters. Part design parameters encompass aspects such as the geometry, materials, orientation, and configuration of the part. Different AM processes require specific process parameters. For example, typical parameters for FFF include print bed temperature, nozzle temperature, print speed, layer height, and infill percentage [44]. The sampling method of data acquisition must also be discussed (e.g., grid sampling, randomized sampling, and adaptive sampling [45]). The data acquisition sampling methods should also be discussed. Along with the raw data statistics, this information ensures that the raw data can be reproduced using similar experimental parameters and sampling methods.

### 3.2.3. Data Handling

**RQ12. Are the data preparation techniques discussed, if any?**

**Explanation:** This question addresses the techniques that use engineering knowledge to prepare and align raw datasets, such as data cleaning, feature extraction, data integration, and data registration. Unlike the ML domain, which primarily focuses on data preprocessing, data preparation plays a vital role in industrial ML applications where data are acquired from multiple complex sources. For instance, time- and frequency-domain features are often extracted from raw time-series data for signal processing, while texture and contour filters in computer vision can be used to extract meaningful features from images [46]. Data from various sources must be registered and integrated at appropriate scales and rates, utilizing domain-specific expertise, before training ML models [47]. If such data preparation methods are employed, they should be detailed to show the domain knowledge required for transforming and aligning the datasets.

**RQ13. Are the data preprocessing techniques discussed, if any?**

**Explanation:** This question addresses the techniques used to preprocess data into a suitable format for ML, aiming to enhance learning performance. Techniques such as scaling, resizing, data splitting, sliding windows, feature selection, and data augmentation fall under this category [45]. This question focuses purely on the data preprocessing techniques from the ML domain. Any techniques applied to preprocess datasets to produce the final input and output data to facilitate ML should be detailed.

**RQ14. Is there a link to a downloadable version of the dataset?**

**Explanation:** Providing a downloadable version of the dataset can significantly ease the effort required for reproduction. Having access to the original dataset allows researchers to use knowledge transfer techniques, which can reduce the number of experiments needed to reproduce the system.

#### 3.2.4. Modeling

**RQ15. Is there a clear description of the machine learning algorithm?**

**Explanation:** This question addresses the foundational aspects of the ML algorithm used in the ML-based IIPQ-AM system, excluding the specific hyperparameters and training methods of the final model. The descriptions of the ML algorithm include the algorithm type (e.g., diffusion models and transformers), type of learning task (e.g., regression, classification, clustering, supervised, semi-supervised, and unsupervised [48]), and specialization (e.g., multifidelity, multiscale, multimodal, and multiobjective [49]). These descriptions provide a fundamental understanding of the ML algorithm, laying the groundwork for a more detailed discussion on the specific trained models used in the system

**RQ16. Are the details of the model described?**

**Explanation:** The structure and architecture of the final model(s) must be clearly presented. For example, in the case of a feedforward neural network (FFNN), details such as the number of hidden layers and the number of neurons in each layer should be provided. Additionally, computational complexity should be discussed. This information is crucial to ensure that readers can accurately reproduce the architecture of the model(s) as proposed in the original paper.

**RQ17. Are the details of model training described?**

**Explanation:** The methods and settings of the ML training procedure must be presented. First, the optimizer must be discussed to reveal which optimization method is chosen, how the learning rate is adjusted, and how the early stopping scheme is scheduled, if applicable. Second, any special ML strategy must be elaborated, including transfer learning, active learning, ensemble learning, and federated learning. Moreover, for debugging and fair comparison, any applicable initial values for pseudorandom number generator (e.g., random seed) can be provided. Similarly, the batching strategy and any balancing or weighting technique used should be mentioned. This question ensures that readers can replicate the training process and achieve the same trained models as those described in the original paper.

**RQ18. If applicable, are hyperparameter search and selection procedures explained?**

**Explanation:** This question concerns how hyperparameters are determined for the selected model(s) (e.g., randomized search or Bayesian optimization). Hyperparameter search and selection is a critical process in ML that may introduce randomness and data leakage, potentially compromising the generalizability of the selected model. Properly addressing this question ensures that the best models are selected from the candidate ML algorithms, which are then compared to demonstrate the advantages of the proposed method. In addition to enhancing trustworthiness, this question helps readers select the same ML model as described in the original paper.

**RQ19. Is the final trained model shared?**

**Explanation:** A downloadable version of the model will significantly increase the success rate of reproduction. By accessing the model, readers can examine the architecture and performance more closely, facilitating a deeper understanding and enabling more accurate reproduction of the results. If multiple files or checkpoints are available, the best-performing should be highlighted.

**RQ20. Is the code open access or provided?**

**Explanation:** A downloadable version of the code will significantly reduce the effort of reproduction. By providing the code, readers can bypass the need to implement the ML architecture, training procedures, and hyperparameter search from scratch, thereby streamlining the reproduction process.

#### 3.2.5. Evaluation

**RQ21. Are the methods and metrics of model selection illustrated?**

**Explanation:** The methods (e.g., cross-validation) and metrics (e.g., accuracy, F1-score) used to evaluate model performance should be discussed. It is essential to demonstrate the suitability of these evaluation methods and metrics in relation to the dataset, ML model, and industrial application. This ensures that readers can apply the same evaluation strategies as described in the original paper, maintaining consistency in the assessment of the model performance.

#### 3.2.6. Deployment

**Special notes on the deployment phase:** The original deployment phase of the CRISP methodology primarily addresses the deployment of the system for business operations, excluding the training infrastructure used during the modeling phase. However, most ML-based IIPQ-AM publications typically demonstrate their proposed systems within laboratory settings, without disclosing the infrastructure necessary for actual business operations. To avoid duplicating questions and imposing unnecessary requirements, the reproducibility questions in the deployment phase investigate the computational resources for training and/or deploying the models. The authors may just report the training infrastructure if their research does not involve deployment in industrial settings.

**RQ22. Is the computational hardware described?**

**Explanation:** The computational resources and infrastructure must be provided to inform the readers of the computational power to train and/or deploy the model(s).

**RQ23. Is the computational software described?**

**Explanation:** The specifications of dependencies such as the packages and virtual environments used to obtain the proposed models must be introduced. The models obtained from different computational environments can be different even if the same ML architecture, training procedure, and hyperparameter search are implemented.

## 4. CASE STUDIES

This section conducts two case studies that implemented the investigation pipeline and filled the reproducibility checklist while reproducing two published ML-based IIPQ-AM researches [50] (Figure 4). First, we reproduced the published system using the proposed pipeline, during which the missing information in the publication was identified using the checklist. Afterward, we contacted the authors of the publications to gather the missing information, with which the system was reproduced again. The checklist was iteratively updated through communications to fulfill all reproducibility requirements. Finally, the effectiveness of the proposed method was analyzed by comparing the performances of the original system, the reproduced system with published information only, and the reproduced system with complete reproducibility information obtained from the original authors.

### 4.1. Case Study I: FFF warpage detection

This case study reproduces an ML-based IIPQ-AM system that detects warping using a digital camera and a CNN model during the FFF process (Figure 5a) [50]. The print is terminated to avoid material waste and machine damage once warping is detected. This system is representative in this domain because 1) FFF is one of the most popular AM processes for recreation, production, and research; 2) vision-based monitoring systems with digital cameras are widely deployed [7]. We reproduced the ML-based IIPQ-AM system based on a Qidi Tech I-Fast FFF 3D printer (Figure 5b) using the instruments and resources available at ADML.

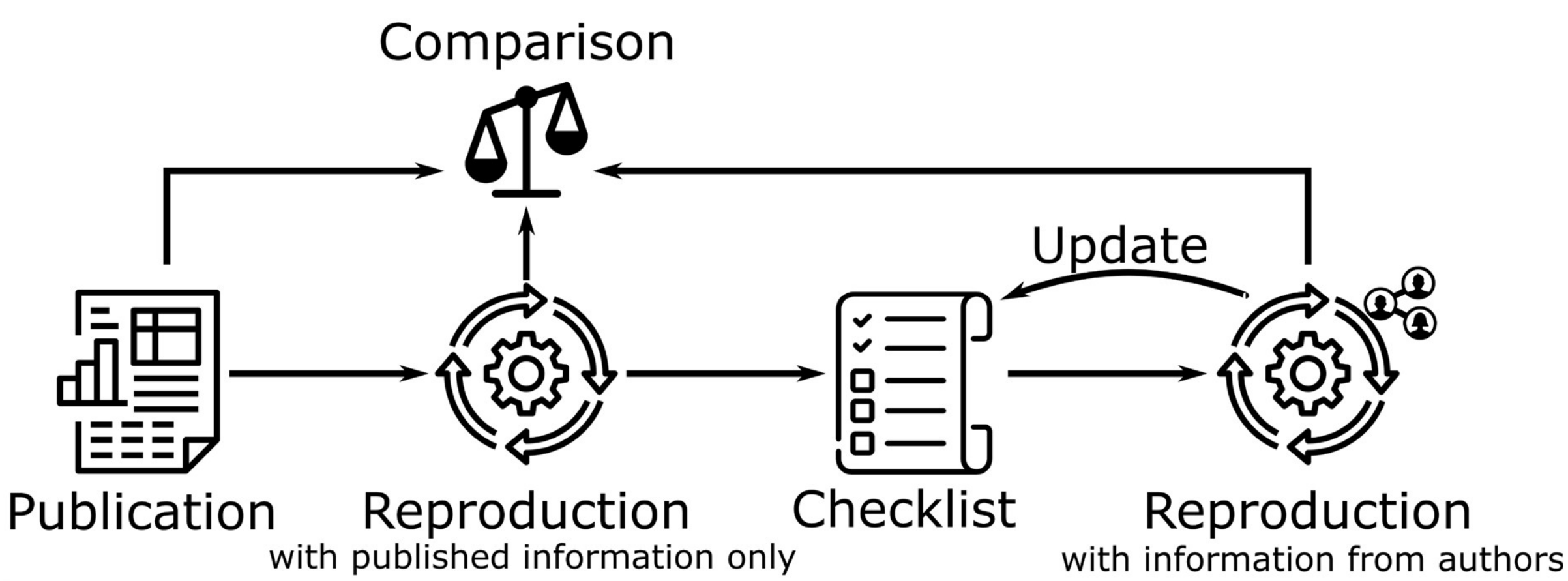

**Figure 4: Implementation pipeline of the case studies.**

Table 2 summarizes the reproduction process of the selected ML-based IIPQ-AM system. We followed the reproducibility investigation pipeline phase-by-phase to replicate the original system, addressing each reproducibility question to evaluate the reproducibility of the original research and develop a reproduction plan. We checked the associated box and recorded the answers if a reproducibility question was answered in the original paper. We left the box empty, recorded the available information, and identified the missing reproducibility information if a question was partially answered or not answered. With the available information, we designed and implemented the reproduced system according to our goals and resources.

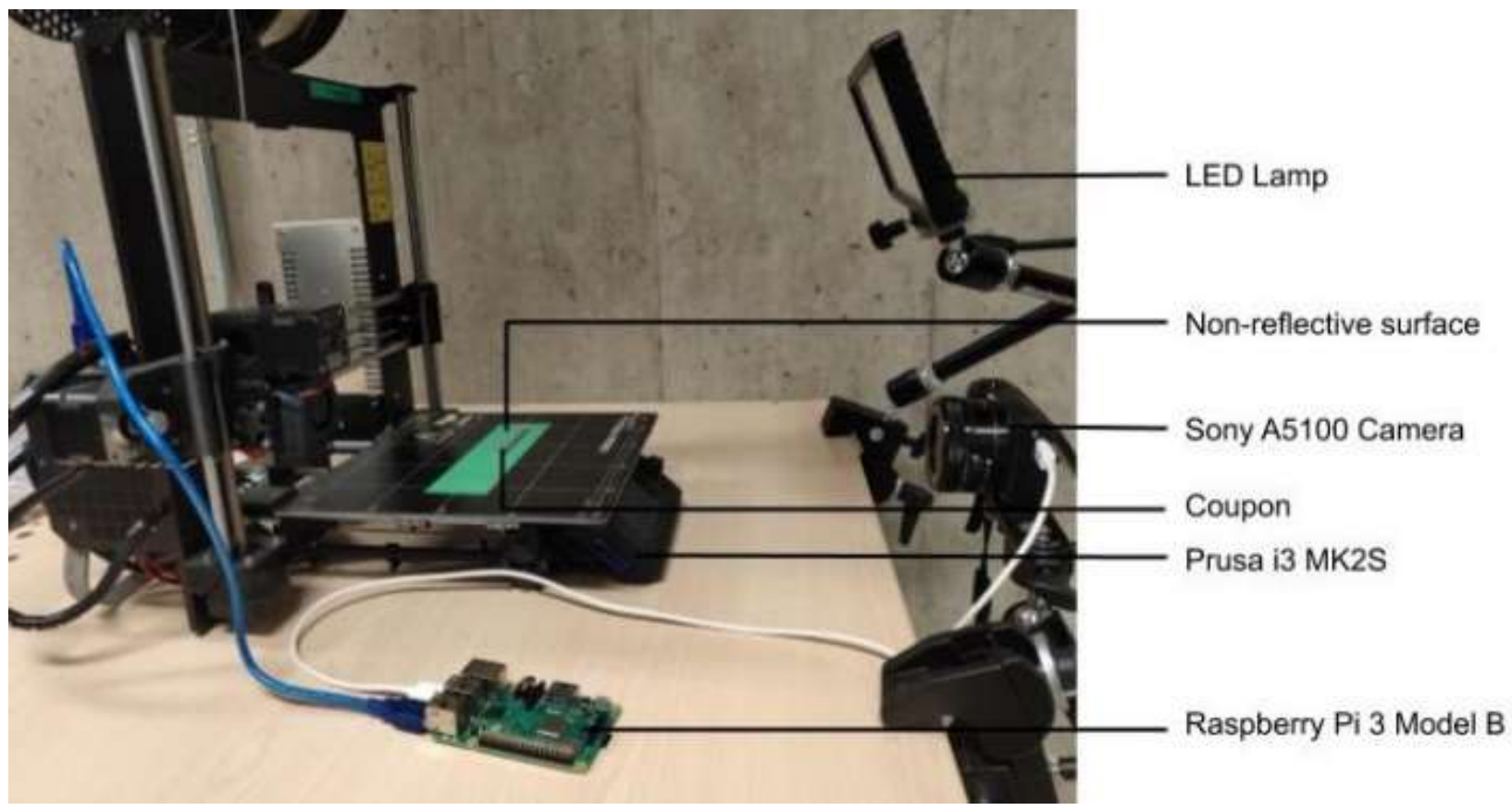

**(a)**

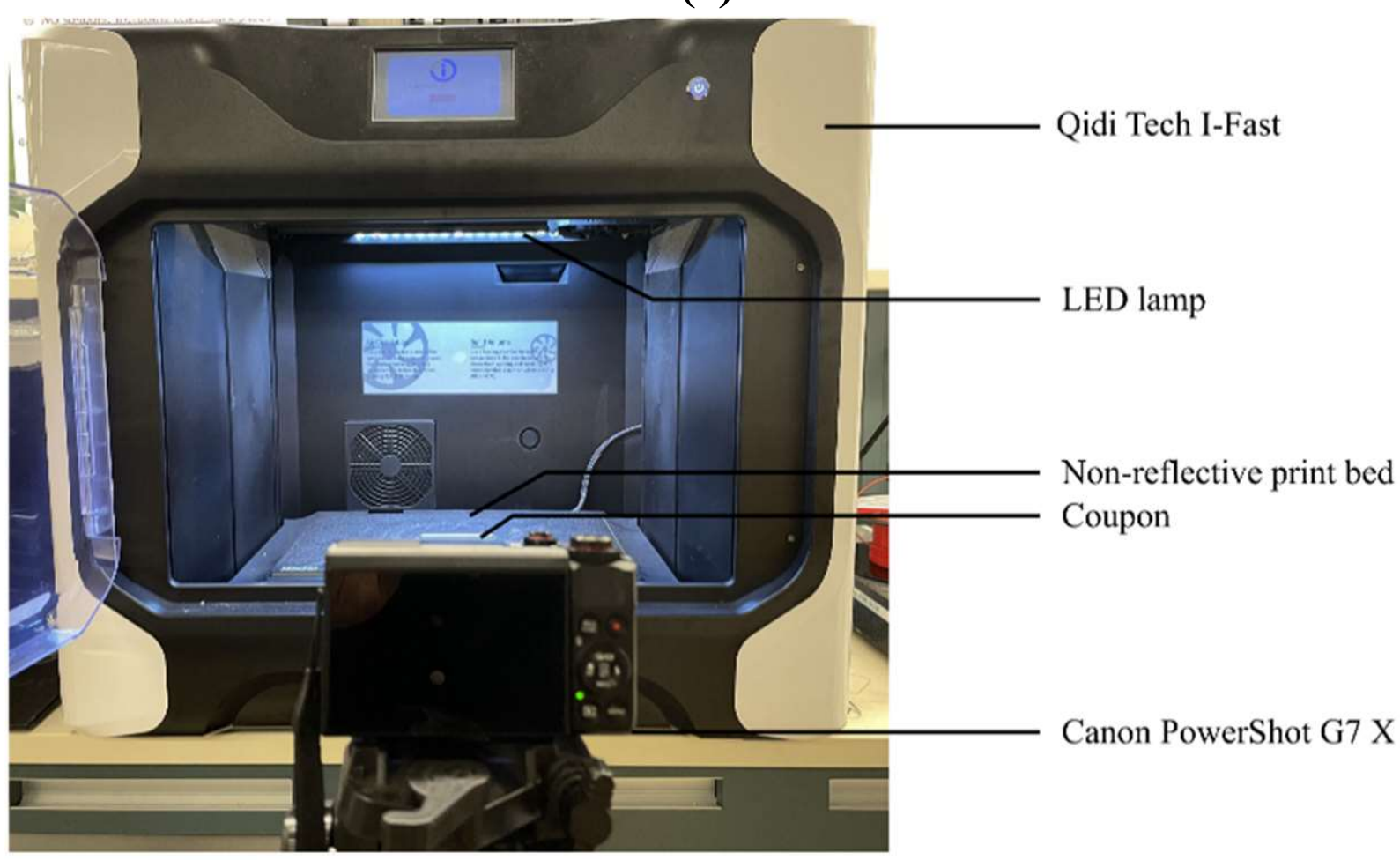

**(b)**

**Figure 5: The experiment setups of a) the original (adapted from [44, 50]) and b) the reproduced ML-based IIPQ-AM systems.**

**Table 2: The implementation of the reproducibility investigation pipeline and the reproducibility checklist for the first case study (FFF warpage detection).**

| ***a) Business Understanding*** |
|---|
| **☒ RQ1. Is the manufacturing system encompassing process technology and associated hardware component(s) reported?**<br>The original AM machine used in the study was an off-the-shelf Prusa i3 MK2S FFF 3D printer. For the reproduction, an off-the-shelf Qidi Tech I-Fast FFF 3D printer was utilized. |
| **☒ RQ2. If applicable, is customization of the manufacturing system reported?**<br>No customization was implemented for the original or reproduced systems. |
| **☒ RQ3. Are characteristics of the material system reported?**<br>The original system used white polylactic acid (PLA) filament with a diameter of 1.75 mm, while the reproduced system used gray PLA filament with the same diameter. |
| **☒ RQ4. Is the machine learning modeling purpose discussed?**<br>The modeling purpose was to detect warping during the FFF process to prevent material waste and machine damage. |
| **☒ RQ5. Are the physical phenomena being captured by the sensing system introduced?** |

The physical phenomenon being monitored was the part geometry.

***b) Data Understanding***

**☒ RQ6. Are sensor type(s) and specification(s) provided?**

The original system used a Sony A5100 digital camera, while the reproduced system used a Canon PowerShot G7 X digital camera.

**☐ RQ7. Are the sensor settings or calibration details discussed?**

The original paper does not discuss the sensor settings or calibration details explicitly. However, the example images provided in the paper suggest some calibration information. No special filters or modes were used, as the images were regular RGB. The only calibration involved adjusting the camera focal length to focus on the part. Accordingly, the camera in the reproduced system was calibrated similarly to focus on the part.

**☒ RQ8. Are the sensor deployment details discussed?**

In the original system, the camera was positioned directly in front of the 3D printer at the same height as the print bed using a desk camera mount (Figure 5a). To enhance image quality, non-reflective tape was placed on the print bed, and an LED lamp was used to increase light intensity and eliminate shadows. For the reproduced system, the camera was mounted at the same height as the print bed using a tripod. To replicate the original setup, a non-reflective print bed and an LED lamp were also used (Figure 5b).

**☒ RQ9. Are the data modalities, formats, and types introduced?**

The raw data for both the original and reproduced systems were RGB images, while the final input data were converted to grayscale images.

**☒ RQ10. Are the relevant statistics provided?**

The original raw dataset consists of 520 RGB images with red, green, and blue channels. The resolution of the raw images is 6000 × 4000 pixels (Figure 6). The dataset has two classes: images of parts without warping and images of parts with warping. There is no data imbalance in this dataset because each class has 260 examples. The final dataset is comprised of 520 grayscale images with one channel and a resolution of 100 × 100 pixels.

**☒ RQ11. Are the experimental design settings for data acquisition introduced?**

In the original experiments, the printing processes of the rectangular coupons were monitored using the experiment setup in Figure 5a. An image was captured at the end of each layer of the printing process. Warping was induced by manually peeling the part from the print bed during the early layers, causing warping angles to increase as thermal stress built up in subsequent layers. The reproduced system followed the same approach for setting up the experiment, creating warping, and capturing layer-wise images. The FFF process parameters used in the original paper were also applied in the reproduced system.

***c) Data Handling***

**☐ RQ12. Are the data preparation techniques discussed, if any?**

The original system did not involve many data preparation techniques. The only noticeable data preparation is cropping the ROI, which is the corner of the part shown in Figure 6. The method for identifying and cropping the ROI was not detailed in the original paper. For the reproduced system, images were cropped manually, positioning the part corners at the center of the cropped images (Figure 6).

**☐ RQ13. Are the data preprocessing techniques discussed, if any?**

The original paper detailed that the raw dataset was shuffled and divided into training and test sets, with 416 and 104 images, respectively. After cropping, the images were resized and converted to grayscale. For labeling, one-hot encoding was used to differentiate between classes. The training set was further split into training and validation subsets, though the exact split ratio was not specified. For the reproduced system, the dataset was prepared following these same procedures, with the split ratio for the training and validation sets adjusted based on typical practices in similar studies.

**☐ RQ14. Is there a link to a downloadable version of the dataset?**

The dataset is not publicly available but is available upon request.

| *d) Modeling* |
|---|
| ☒ **RQ15. Is there a clear description of the machine learning algorithm?**<br>The selected ML algorithm was CNN. The original paper provided a detailed description of the CNN model structure, specifying the convolutional layers, pooling layers, fully connected layers, and activation functions used in the model. |
| ☐ **RQ16. Are the details of the model described?**<br>The original paper clearly exhibited most of the hyperparameters of the final CNN model, including the activation functions, dropout rate, number of hidden layers, number of neurons in each layer, kernel sizes, stride, and activation functions. However, the initial learning rate was not specified. |
| ☒ **RQ17. Are the details of model training described?**<br>Adam optimizer was used to train the CNN model for 50 epochs, with an early stopping mechanism implemented to terminate the training process when the validation loss converged. The same training method was applied to the reproduced system. |
| ☐ **RQ18. If applicable, are hyperparameter search and selection procedures explained?**<br>The original paper did not discuss the hyperparameter search and selection method. For the reproduced system, a self-defined hyperparameter optimization approach was implemented. |
| ☐ **RQ19. Is the final trained model shared?**<br>No. |
| ☐ **RQ20. Is the code open access or provided?**<br>No. |
| *e) Evaluation* |
| ☒ **RQ21. Are the methods and metrics of model selection illustrated?**<br>The original paper used classification accuracy as the metric to evaluate model performance, and this metric was also applied to the reproduced system. Since the original paper did not discuss hyperparameter search, a model selection method was not applicable. |
| *f) Deployment* |
| ☒ **RQ22. Is the computational hardware described?**<br>For both the original and the reproduced systems, the training and predictions of the networks were run in an Intel® Core™ i5-6300U CPU @ 2.40 GHz and 8GB memory running on Windows 10 64-bits. The Raspberry Pi in the original system controlled the camera to automatically take a picture once a layer was completed. In the reproduced system, the camera automatically took a picture at every minute. |
| ☐ **RQ23. Is the computational software described?**<br>The only information about the software was that the programming language was Python 3.6. |

The reproduction of the selected ML-based IIPQ-AM system initially relied solely on the information available from the original paper. We reproduced the manufacturing and sensing systems as described in the business understanding and the data understanding phases (Table 2 and Figure 5). The only missing information regarding the camera settings and calibration was inferred from the example images in the original paper. Experiments were then conducted with the reproduced system based on the original experimental design (Figure 6). Fifteen coupons were sequentially printed, each featuring one corner with warping and another without warping. Each coupon consisted of 20 layers and had a print time of approximately 20 minutes. Images were captured starting from the 6th minute of each print, resulting in 214 images with warping and 204 images without warping after removing outliers. During the data handling phase, the data preparation and preprocessing techniques discussed in the original paper were applied. For the missing information, we cropped the ROIs with the part corners at the center and assigned an 8:2 ratio to the train-validation split. The ML models of the reproduced system were trained and evaluated according to the original paper.

The remaining missing information pertained to the initial learning rate and the hyperparameter search method. We trained the candidate models in two ways: 1) using the original hyperparameters and an initial learning rate of 0.01(Figure 7a) and 2) conducting a randomized hyperparameter search with self-defined hyperparameter ranges (Figure 7b). The first approach achieved validation and test accuracies of 92.6% and 92.8%, respectively. The second approach yielded an optimal model with validation and test accuracies of 95.1% and 94.9%, respectively. These results indicate that the original hyperparameters were not optimal for the reproduced system. The changes in the manufacturing and sensing systems imposed a domain shift on the distribution of the input image data. A hyperparameter search for the reproduced system can address this by finding the new optimal hyperparameters. We conducted a self-defined hyperparameter search because the original hyperparameter search information was missing. However, if the original hyperparameter search information had been available, it could have served as a valuable reference to guide the selection of hyperparameters and the ranges for the search, potentially increasing efficiency and improving performance.

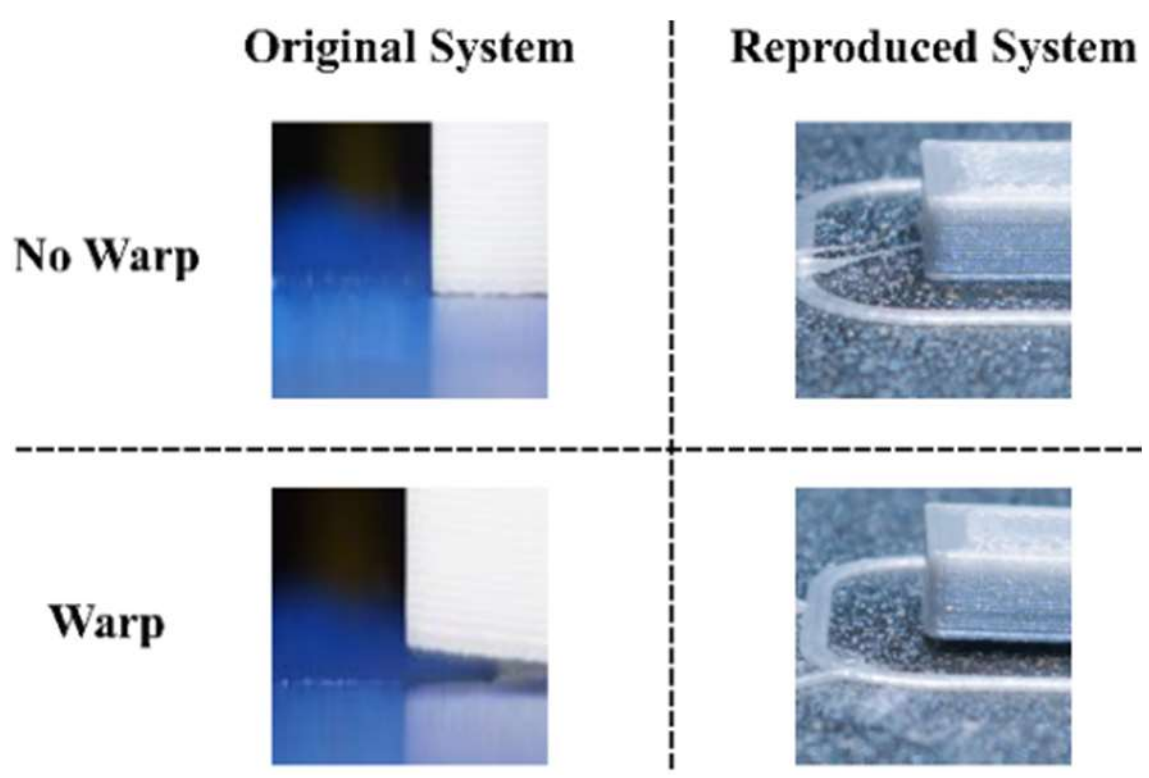

**Figure 6: Examples of original and reproduced cropped images.**

The system was reproduced again after gathering all the reproducibility information from the original authors (Figure 7c). It was confirmed that the original experiments did not involve any special camera settings or calibration, except for adjusting the focal length to focus on the part. One detail not disclosed in the original paper was the use of non-reflective tapes of different colors to enhance the representativeness of the dataset. Additionally, the original manual cropping method only required the part corners to be included in the cropped images, rather than centered, to improve the robustness of the model. This adjustment was necessary because, in practice, the corners cannot always be centered in the cropped image due to variations in part positions and geometries. To approximate these effects in the reproduced system, data augmentation was applied to the dataset. The original train-validation split ratio of 8:2 was also implemented. Finally, the same hyperparameter search as in the original system was conducted, including the search method (Bayesian optimization), tunable hyperparameters, and search ranges. The optimal model achieved validation and test accuracies of 98.1% and 98.4%, respectively.

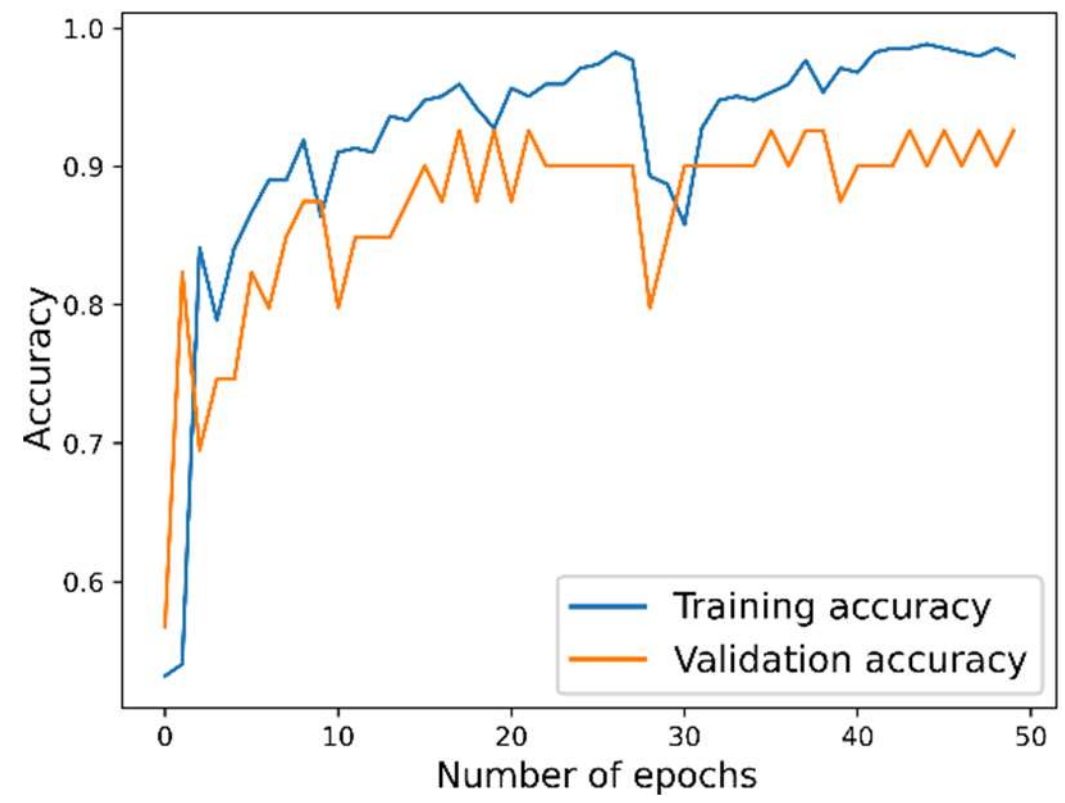

**(a)**

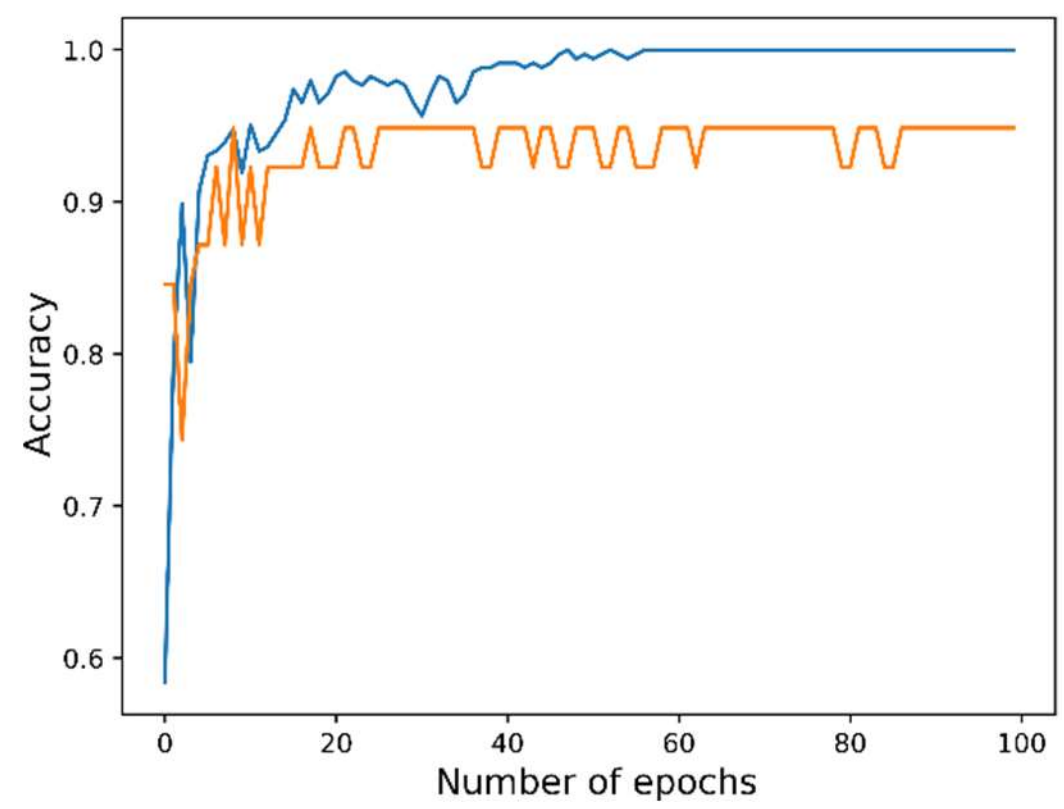

**(b)**

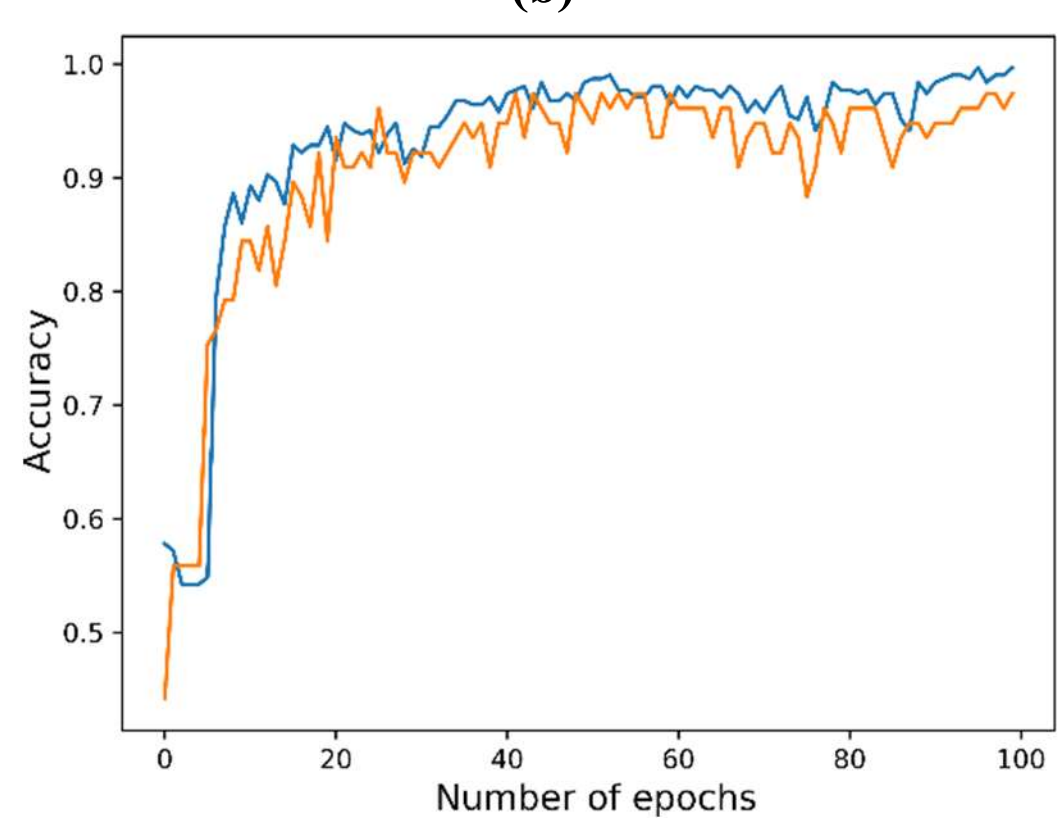

**(c)**

**Figure 7: Accuracy curves of the reproduced CNN model when using a) the information available from the original paper; b) the information available from the original paper and a self-defined hyperparameter search; and c) complete reproducibility information provided by the original authors.**

By obtaining and adopting the complete reproducibility information, the test performance of the reproduced system increased from 94.9% to 98.4%, bringing it much closer to the 99.3% reported in the original paper. With the original hyperparameter search information, the search process became more efficient. The newly added experimental design and data handling details significantly enhanced the generalizability of the model. These improvements reduced overfitting, resulting in smaller discrepancies among the training, validation, and test accuracies. Based on the first case study, we identified the following key takeaways from reproducing vision-based IIPQ-AM systems for FFF:

- Layer-wise images are subject to variations such as different filament colors, print bed surfaces, and background interferences. These variations often need to be addressed in the reproduced system through experimental design or data augmentation.
- Different manufacturing and sensing equipment in the original and reproduced systems lead to domain shifts. As a result, the optimal hyperparameters for the reproduced system will differ from those in the original system, making hyperparameter search a necessary step during the reproduction process.

### 4.2. CASE Study II: LPBF melt pool area prediction

The second case study aims to reproduce an in-process melt pool area prediction model for LPBF, utilizing the raw data collected from the additive manufacturing metrology testbed (AMMT) at the National Institute of Standards and Technology (NIST) [51]. Vision-based monitoring systems that extract morphological and thermal characteristics of melt pools can detect process anomalies and predict part quality. Ko et al. [51] trained an FFNN model to predict the melt pool area based on the processing parameters and the melt pool neighboring effects. This case study represents a large number of publications that report melt pool monitoring systems for LAM. In [51], the authors utilized the dataset from their previous research instead of acquiring data from new experiments. Usually, the papers utilizing previously disclosed datasets only offer concise information about the experimental design and apparatus, while pasting the references for readers to explore the details. Thus, we consider the reproducibility information to be present if the reproducibility information can be obtained from the references. Similar to the original publication [51], this case study only trained the melt pool area prediction model based on the existing dataset without reproducing the manufacturing and sensing systems.

The raw dataset, including the process parameters and melt pool images, was shared by the original authors upon request. The main data preparation technique in the original paper was the neighboring-effect modeling (NBEM) which models the spatiotemporal relationships among the registered melt pools. We implemented NBEM and derived physics-informed features including NBEM time and distance. The raw images and final dataset are shown in Figure 8. Table 3 records the answers to the reproducibility questions while investigating and reproducing the model presented in [51]. The reproducibility checklist was filled out using the same principles as the first case study.

Table 3 indicates that the original paper missed reproducibility information regarding the data preprocessing, model structure, model training, and hyperparameter search. The initial reproduction filled the information gap by implementing no data preprocessing or hyperparameter search technique and adopting the most common model structure and training techniques. For example, the initial model adopts an activation function of rectified linear unit (ReLU), an initial learning rate of 0.001, a loss function of mean squared error (MSE), and a total of 300 training epochs. We trained the model using PyTorch in Python, which prevails in the ML community. The cross-validation mean absolute percentage error (MAPE) achieved by the initial reproduction was 2036.59%, which was considerably larger than the MAPE achieved in the original paper (15.23%).

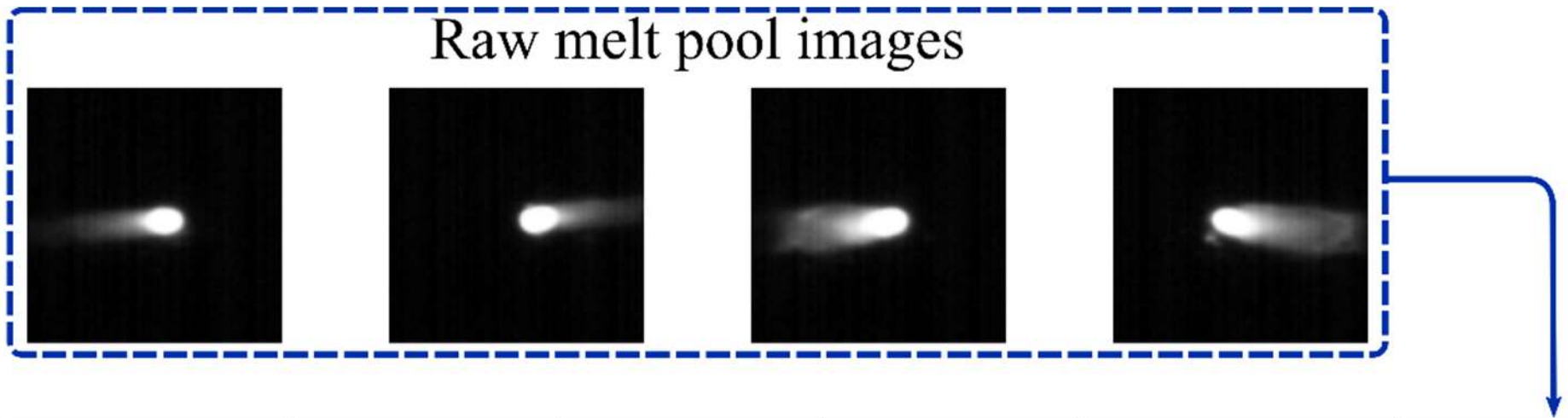

| Building Time | Laser Power | Scan Speed | NBEM Time | NBEM Distance | Melt Pool Area |
|---|---|---|---|---|---|
| 98501 | 104.7 | 902 | 0 | 0 | 0.006208 |
| 1.08E+05 | 147.04 | 100 | 0 | 0 | 0.01248 |
| 1.09E+05 | 194.89 | 468 | 0 | 0 | 0.01152 |
| 1.09E+05 | 232.33 | 803 | 1.55E-05 | 0.58739 | 0.01824 |
| 1.10E+05 | 234 | 803 | 21.949 | 7.1661 | 0.017408 |
| 1.10E+05 | 234 | 803 | 11.685 | 4.4687 | 0.02368 |
| 1.11E+05 | 234 | 767 | 0 | 0 | 0.020864 |
| ... | ... | ... | ... | ... | ... |

**Figure 8: Raw melt pool images and the final tabular dataset.**

**Table 3: The implementation of the reproducibility investigation pipeline and the reproducibility checklist for the second case study (LPBF melt pool area prediction).**

| |
|---|
| ***a) Business Understanding*** |
| **☒ RQ1. Is the manufacturing system encompassing process technology and associated hardware component(s) reported?**<br>The AM system was AMMT from NIST. |
| **☒ RQ2. If applicable, is customization of the manufacturing system reported?**<br>The customizations of AMMT, including the energy, controller, material, and environment control subsystems, are specified in [52]. |
| **☒ RQ3. Are characteristics of the material system reported?**<br>The original system used Inconel-625 powders. |
| **☒ RQ4. Is the machine learning modeling purpose discussed?**<br>The modeling purpose was to predict the melt pool area based on the process parameters and melt pool images captured by a high-speed camera. |
| **☒ RQ5. Are the physical phenomena being captured by the sensing system introduced?**<br>The physical phenomenon being monitored was the melt pool geometry and the correlation between the process parameters and the melt pool geometry. |
| ***b) Data Understanding*** |
| **☒ RQ6. Are sensor type(s) and specification(s) provided?**<br>The original system employed a co-axial Mikrotron EOSens 3CL camera [53, 54]. |
| **☒ RQ7. Are the sensor settings or calibration details discussed?**<br>The sensor settings or calibration details were clearly described in the references [53, 54]. For example, the pixel pitch was 8 µm, the window size was 120 × 120 pixels, the magnification was 1×, the frame rate was 10,000 Hz, and the shutter speed was 20 µs. |
| **☒ RQ8. Are the sensor deployment details discussed?**<br>The deployment details of the vision-based monitoring system of AMMT are illustrated in [53, 54]. |
| **☒ RQ9. Are the data modalities, formats, and types introduced?** |
| The raw data contained tabular data and grayscale images. The final input dataset was tabular data. |
| ☒ **RQ10. Are the relevant statistics provided?**<br>The raw dataset consists of 4700 grayscale images. The resolution of the raw images is 120 × 120 pixels (Figure 8). Each melt pool image was labeled with the melt pool area extracted from the in-process melt pool images. The final tabular dataset contains 4700 rows and 6 columns, consisting of five input features (building time, laser power, scan speed, NBEM time, and NBEM distance) and one label (melt pool area). |
| ☒ **RQ11. Are the experimental design settings for data acquisition introduced?**<br>A laser power of 100 W was used with a scan speed of 900 mm/s for the precontour, while the infill hatching was performed at 195 W and 800 mm/s. The part is made up of 250 layers using an island-spiral-concentrating scan pattern, with a rotation angle of 67° between layers. |
| ***c) Data Handling*** |
| ☒ **RQ12. Are the data preparation techniques discussed, if any?**<br>The authors conducted data registration for the raw data, spatially and temporally aligning the melt pool images [36]. Features such as average and maximum melt pool areas were extracted from the melt pool images. NBEM features were extracted from the registered data to capture the interaction between the melt pools. The same data preparation techniques were implemented in the reproduction. |
| ☐ **RQ13. Are the data preprocessing techniques discussed, if any?**<br>The original paper did not mention any data preprocessing techniques. However, an inquiry revealed that Ko et al. [51] applied normalization to the raw dataset. |
| ☐ **RQ14. Is there a link to a downloadable version of the dataset?**<br>The dataset is not publicly available but is available upon request. |
| ***d) Modeling*** |
| ☒ **RQ15. Is there a clear description of the machine learning algorithm?**<br>The selected ML algorithm was FFNN. |
| ☐ **RQ16. Are the details of the model described?**<br>The original paper provided the number of hidden layers and number of neurons of the FFNN model. However, other crucial information such as the activation function and initial learning rate was not discussed. |
| ☐ **RQ17. Are the details of model training described?**<br>The authors used Levenberg-Marquardt training method and leave-one-out cross-validation to train the model. However, some key training information such as the loss function and number of epochs was not provided. |
| ☐ **RQ18. If applicable, are hyperparameter search and selection procedures explained?**<br>Hyperparameter search was not discussed in the original paper. |
| ☐ **RQ19. Is the final trained model shared?**<br>No. |
| ☐ **RQ20. Is the code open access or provided?**<br>No. |
| ***e) Evaluation*** |
| ☒ **RQ21. Are the methods and metrics of model selection illustrated?**<br>The method to test the model was leave-one-out cross-validation. The model evaluation metric was MAPE. |
| ***f) Deployment*** |
| ☐ **RQ22. Is the computational hardware described?**<br>No. |
| ☐ **RQ23. Is the computational software described?**<br>No. an inquiry revealed that Ko et al. [51] used MATLAB and the fitnet function. |

After obtaining the missing information from the original authors, we reproduced the model again: 1) normalization was conducted to scale the raw dataset; 2) early stopping was implemented to terminate the training process once the training converged; 3) the model used ReLU as the activation function, MSE as the loss function, and an initial learning rate of 0.001. Although the original authors revealed that they trained the model using the MATLAB fitnet function, we still used Python and replicated the settings of the fitnet function to be consistent. The cross-validation MAPE was successfully reduced to 15.13% with complete reproducibility information. Figure 9 visualizes the cross-validation residuals of the reproduced model using a histogram with a smooth and continuous blue line representing the estimated probability density function generated by kernel density estimate (KDE). The histogram shows a bell-shaped distribution of residuals centered around zero, with a standard deviation of 0.0029. The tightly clustered residuals suggested that the predictions were subject to small bias and variance.

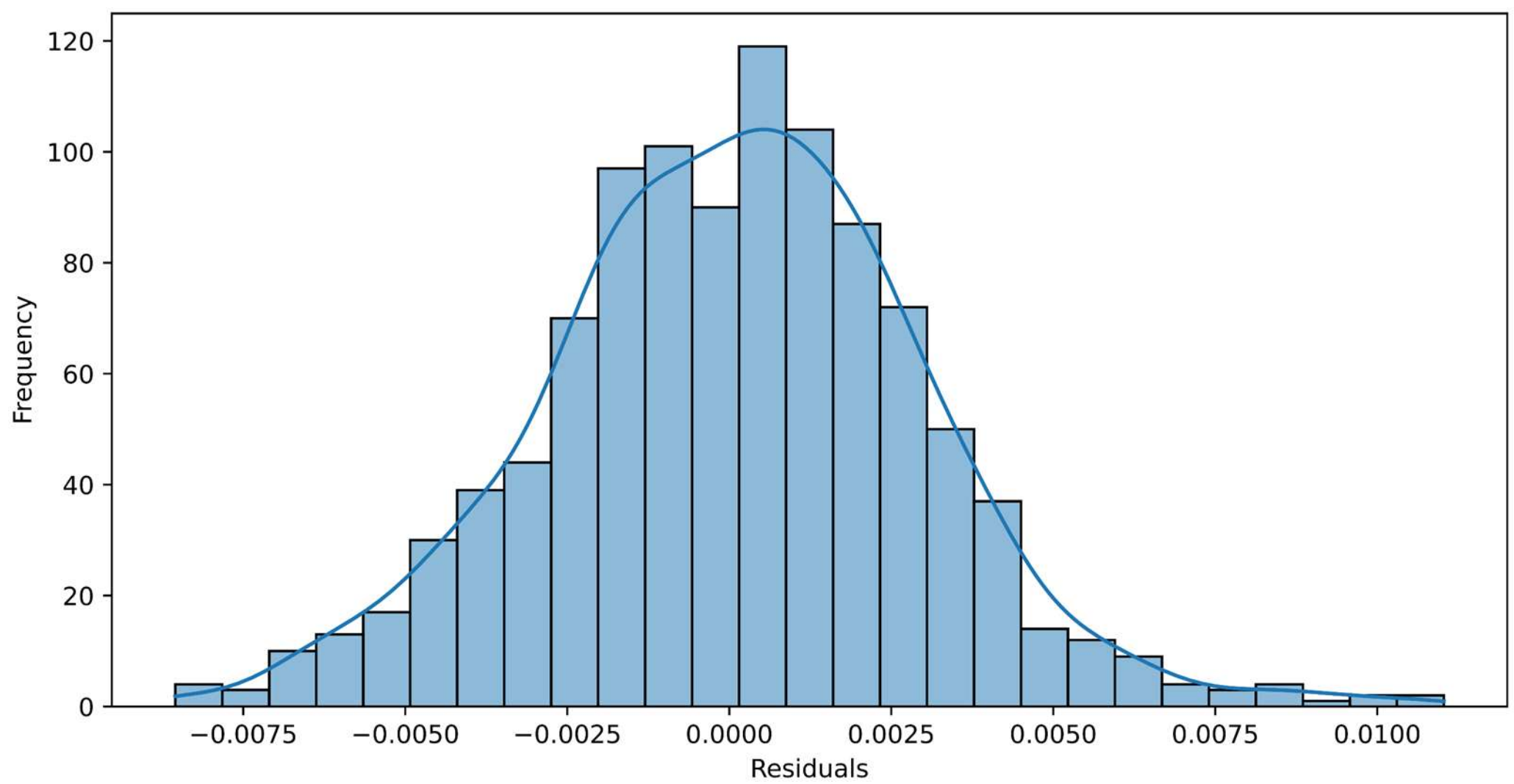

**Figure 9: Histogram of the cross-validation residuals of the reproduced melt pool area prediction model.**

Based on the second case study, we identified the following key takeaways from reproducing a melt pool area prediction model for LPBF:

- Differences in the performance of models trained using different ML libraries, such as PyTorch and fitnet, can arise even when the same network structures are employed. This variation is often due to the underlying optimizers and schedulers used by these libraries, which may have distinct default parameters.
- Details of the model structure and training methods are crucial for reproducing an ML model. Differences in a few key ML settings and hyperparameters can lead to significant performance differences.

The case studies demonstrated that the proposed method could enhance the performance of reproduced ML-based IIPQ-AM systems. The reproducibility checklist ensures that the authors provide all the reproducibility information in their publications. The investigation pipeline helps the readers investigate the reproducibility of an ML-based IIPQ-AM research and guides them through the reproduction.

## 5. REPRODUCIBILITY SURVEY

We conducted a reproducibility survey for ML-based IIPQ-AM publications based on the reproducibility checklist to investigate the level of reproducibility in this domain. The SciHIT IE platform was utilized to facilitate the extraction of reproducibility information from the publications.

### 5.1. Survey methodology

This survey utilized the Scopus database to collect relevant publications, focusing on journal papers published between 2018 and 2023. We employed the search criteria in Table 4 to search the publication titles, abstracts, and keywords. Major terminologies that describe AM processes are included in the AM keywords. "Machine learning" and "deep learning" are included as the ML keywords to filter in ML-based monitoring systems. Phrases related to process monitoring are also added as application keywords. The AM, ML, and applications keyword groups are connected using "AND". Subject areas other than engineering, computer science, and material science are excluded.

**Table 4: Search criteria for searching the title, abstract, and keywords using Scopus.**

| Additive manufacturing | Machine learning | Application |
|---|---|---|
| "Additive Manufacturing" OR "3D Printing" OR "Filament Fused Fabrication" OR "Powder Bed Fusion" OR "Directed Energy Deposition" OR "Fused Deposition Modeling" OR "Stereolithography" OR "Binder Jetting" OR "Sheet Lamination" | "Machine Learning" OR "Deep Learning" | "Monitoring" OR "Defect Detection" OR "Anomaly Detection" |

The trend of the relevant publications presented in Figure 10 indicates that the research interest in ML-based IIPQ-AM has been rapidly increasing. However, none of the existing research investigated the reproducibility of the published works. This survey randomly sampled a portion of publications and extracted the reproducibility information from them (Figure 10). The number of papers selected from each year must exceed five and account for at least one-third of the yearly publications to be representative.

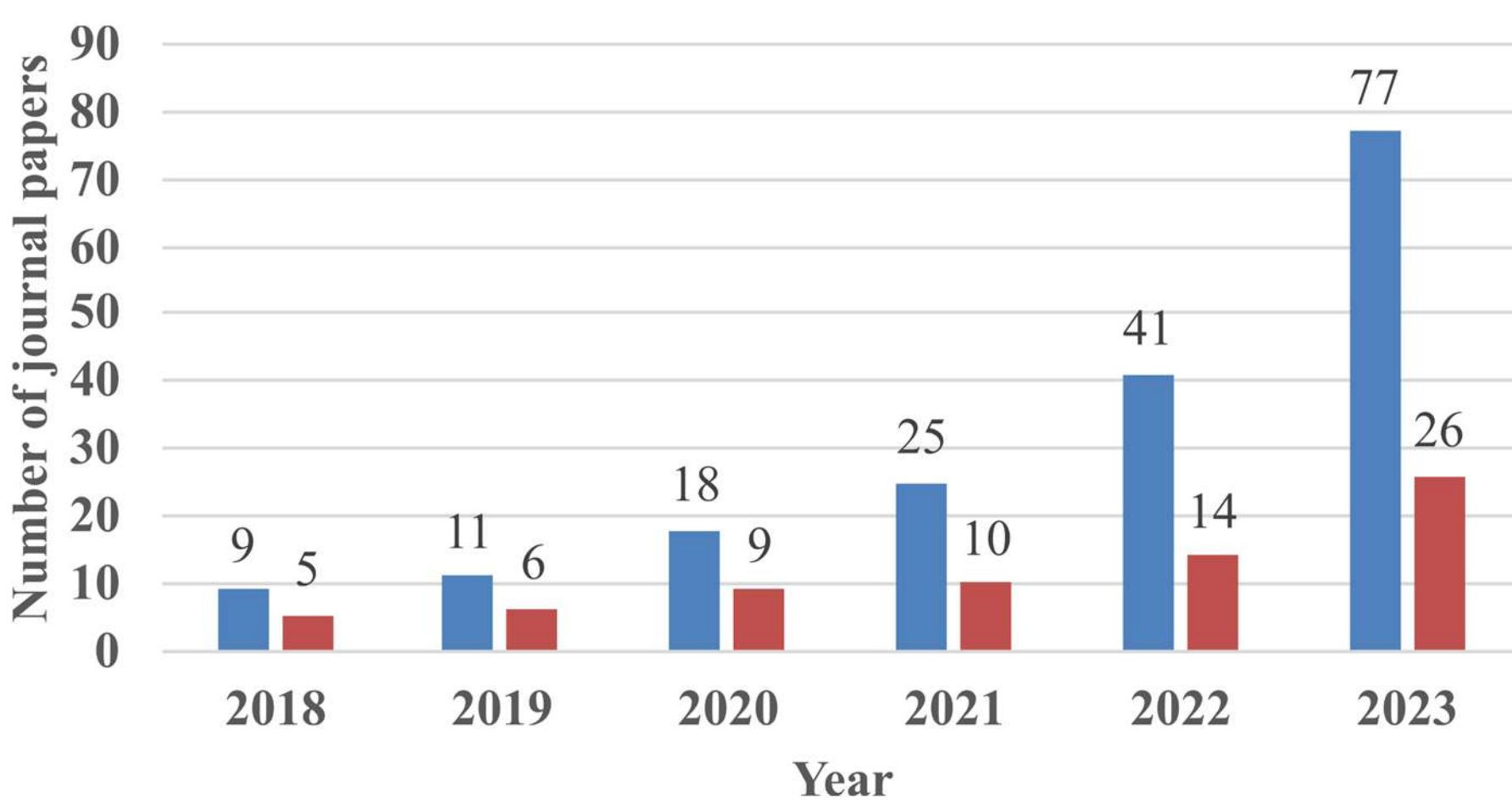

**Figure 10: Number of papers collected based on the search criteria and number of papers surveyed.**

This survey utilized the SciHIT IE platform to facilitate the reproducibility information extraction (IE) from the selected publications. The SciHIT IE platform developed by ADML is a cloud-based IE platform accessible via web browsers [55]. Powered by LLMs and a human-AI teaming framework, the SciHIT IE platform can achieve accurate semi-automatic IE from scientific publications. Figure 11a exhibits the main view of the SciHIT IE platform, consisting of a library (left), an IE interface (upper right), and a publication viewer (lower right). The users can create multiple libraries and upload papers, which are automatically parsed and stored in the database for subsequent IE. Once a publication is selected from the library, the publication viewer displays the original PDF or the parsed texts according to the user preference. The IE interface encompasses multiple IE functionalities:

1. ***Text search***: Traditional text search which compares and matches strings letter-by-letter.
2. ***Semantic search***: Computing cosine similarity between two text embeddings.
3. ***Query***: Creating a prompt to feed the query into a GPT model.
4. ***Retrieval***: Customized IE with self-improving functionality powered by active learning.
5. ***Checklist***: An extension designed for the reproducibility survey. The extracted reproducibility information is populated and stored in the database using this functionality.

The ScieHIT IE platform is established based on a human-AI teaming framework comprised of the base IE system, the paragraph classification tier, and the query tier. (Figure 11b). This framework emphasizes transparency and domain knowledge, improving IE performance in a human-in-the-loop manner. The base IE system manages the uploaded documents and parses the texts and queries, which are compared based on their similarities. The manually uploaded PDFs are listed in the selected library and automatically parsed into texts. The users can browse the documents using either the PDF viewer or the text viewer. According to the IE goals, the user compiles a list of pre-set queries, which are the reproducibility questions in this survey. Both the queries and parsed paragraphs from the documents are converted to vectors using the OpenAI text embedding-ada-002 model. Thereafter, *Semantic search* scores the similarity between the query vectors and the document paragraph vectors using cosine similarity. The paragraphs that are most similar to a query will be highlighted in the text viewer and labeled as relevant paragraphs. AM experts can read the highlighted paragraphs and add additional information to the queries accordingly to improve the accuracy of IE. The paragraph classification tier trains a paragraph classification model based on paragraphs labeled by AM experts to identify the relevant paragraphs. In the query tier, AM experts set up more specific queries to extract answers to the reproducibility questions from the most relevant paragraphs. A prompt embracing the relevant paragraphs and queries is compiled and fed into an OpenAI GPT model, which outputs the answers.

We adapted the SciHIT IE framework to the reproducibility survey by designing the pre-set queries that describe the reproducibility questions. For each paper, we extracted reproducibility information question by question. The reproducibility questions, along with their explanations, were included in the pre-set queries. More detailed descriptions and examples were added to the queries to increase the accuracy of highlighting relevant paragraphs. We skipped the paragraph classification tier and considered the highlighted paragraphs the most relevant paragraphs, from which the *Query* functionality extracted reproducibility information for each reproducibility question. A manual search was conducted if the *Query* output did not sufficiently answer the reproducibility question. According to the reproducibility information extracted, we assigned one of the four criteria to each reproducibility question for each paper:

- **Present**: The reproducibility question is answered with sufficient information.

- **Partial**: The reproducibility question is answered with partial information. Some crucial reproducibility information is missing.
- **Missing**: The paper offers no information regarding a reproducibility question.
- **NA**: The reproducibility question does not apply to this paper. For example, NA is assigned to RQ2 when there is no customization in the manufacturing system.

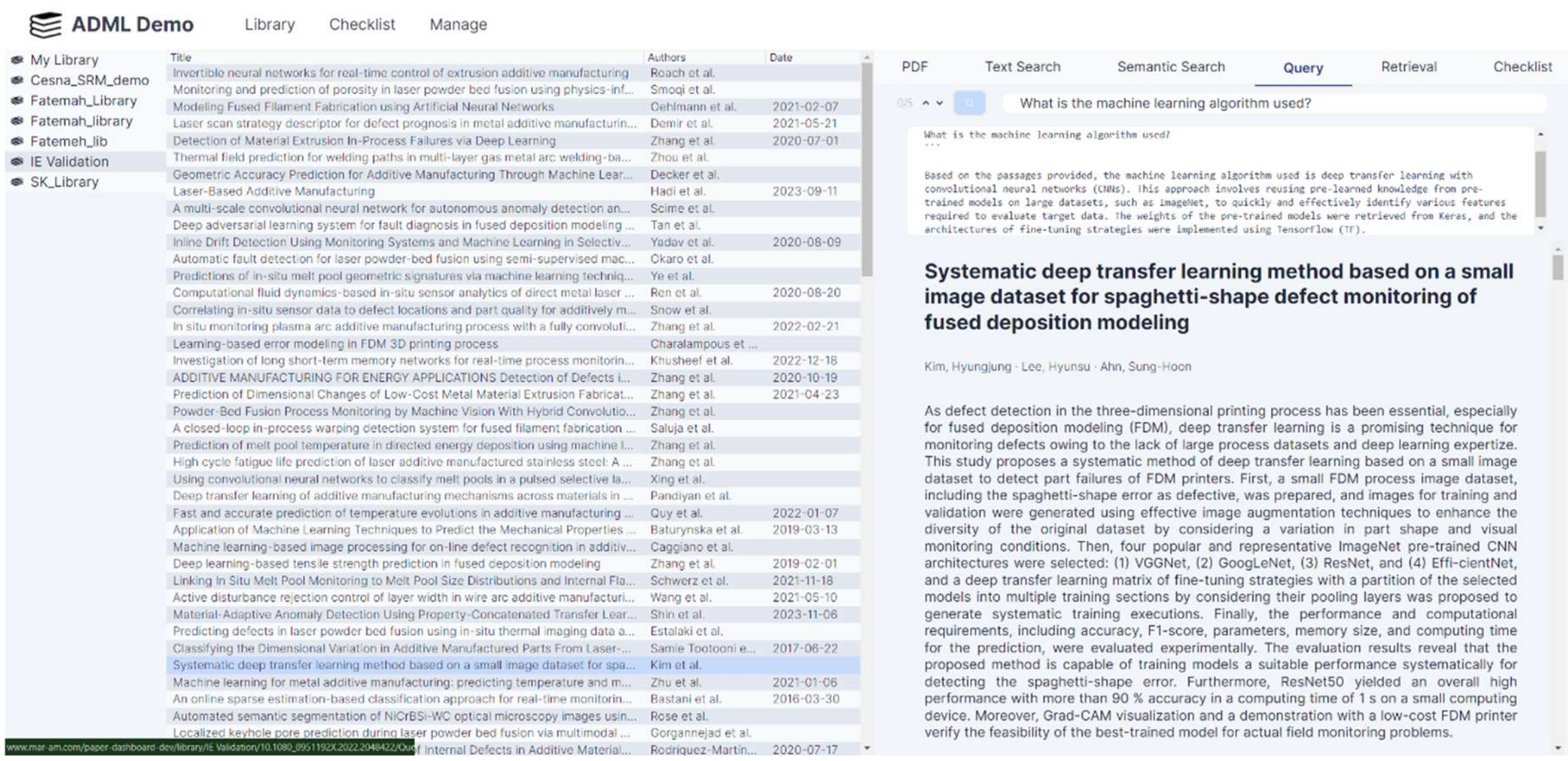

**(a)**

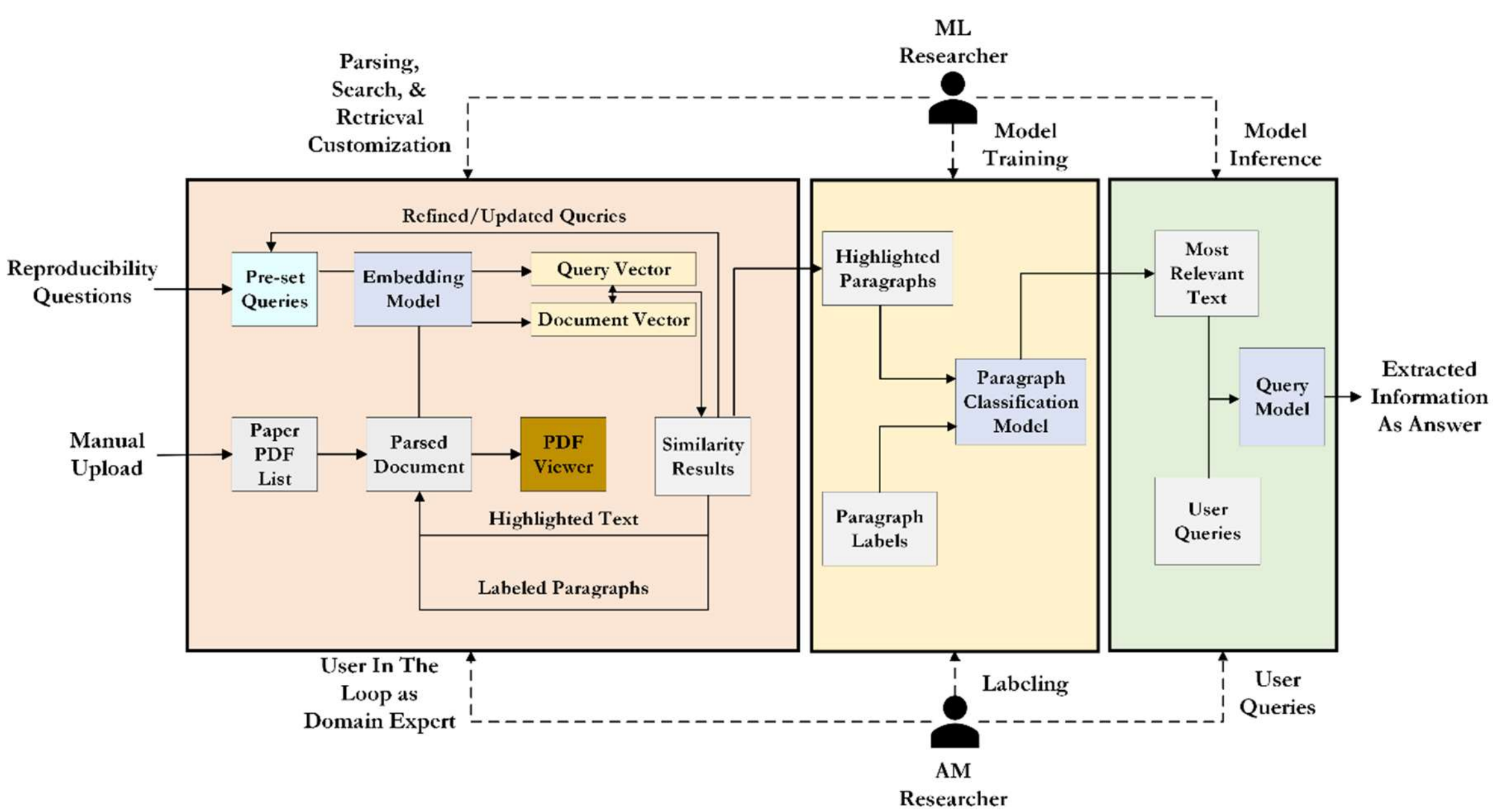

**(b)**

**Figure 11: Schematics of the SciHIT IE platform: (a) web-based platform and (b) human-AI teaming framework of SciHIT IE, including the base IE system (left), paragraph classification tier (middle), and query tier (right) (adapted from [55]).**

## 5.2. Survey results

The survey results reveal the reproducibility status in the ML-based IIPQ-AM domain, which is categorized into manufacturing system-related, sensing system-related, dataset-related, and model-related (Figure 12). This subsection discusses the survey findings according to the four categories. For most questions, the larger the proportion covered by ‘Present’, the more reproducible the research is. Some reproducibility questions correspond to additional information that facilitates the reproduction, including the data, model, and code availabilities (RQ14, RQ19, and RQ20) and the computational hardware and software (RQ22 and RQ23). A lack of additional information might not compromise reproducibility.

Manufacturing system-related reproducibility information is the most complete among all categories, with around 60 out of 70 papers sufficiently providing information regarding the manufacturing machine, customization, and material system. The authors mostly come from manufacturing and industrial domains and possess professional backgrounds in describing manufacturing systems. Besides, the manufacturing systems usually employ off-the-shelf machines and materials, simplifying the reproducibility information. The modeling purposes and physical phenomena (RQ4 and RQ5) are clearly stated in all surveyed papers. Nonetheless, only 55 papers sufficiently reveal the sensor types and specifications (RQ6). Less than 50 papers sufficiently provide the sensor setting, calibration, and deployment details (RQ7 and RQ8). Sensor setting, calibration, and deployment determine the characteristics of the acquired data and thus significantly influence the reproducibility.

Most authors are aware of data reproducibility as more than 60 papers provide at least partial information about the datasets. However, fewer than 55 papers sufficiently describe the basic dataset information (RQ9), dataset statistics (RQ10), and experimental design (RQ11). 60 papers provided at least partial information about data preparation (RQ12) and data preprocessing (RQ13). However, fewer papers offer sufficient data preprocessing information than data preparation due to a lack of ML expertise. Only eight papers have made the dataset publicly available (RQ14) at the time of publication because most datasets are confidential.

The model reproducibility is the worst among all categories with notable missing information. 67 papers clearly described the ML algorithms involved (RQ15). Around 60 papers contain at least partial information about the model structure and training. However, only 46 papers and 39 papers provide sufficient information about the model structure and training (RQ16 and RQ17), respectively. Besides, hyperparameter search (RQ18) is not discussed in 42 papers and 21 papers contain partial information about hyperparameter search. It can be observed that a lack of ML expertise has led to serious model reproducibility problems in this domain. Additionally, code and model availabilities (RQ19 and RQ20) are low because they are mostly confidential. Computational software and hardware (RQ22 and RQ23) deployed in the papers are not commonly discussed, which might not significantly deteriorate the model reproducibility. 55 papers sufficiently provide the model evaluation (RQ21) information, whereas four papers do not provide relevant information at all.

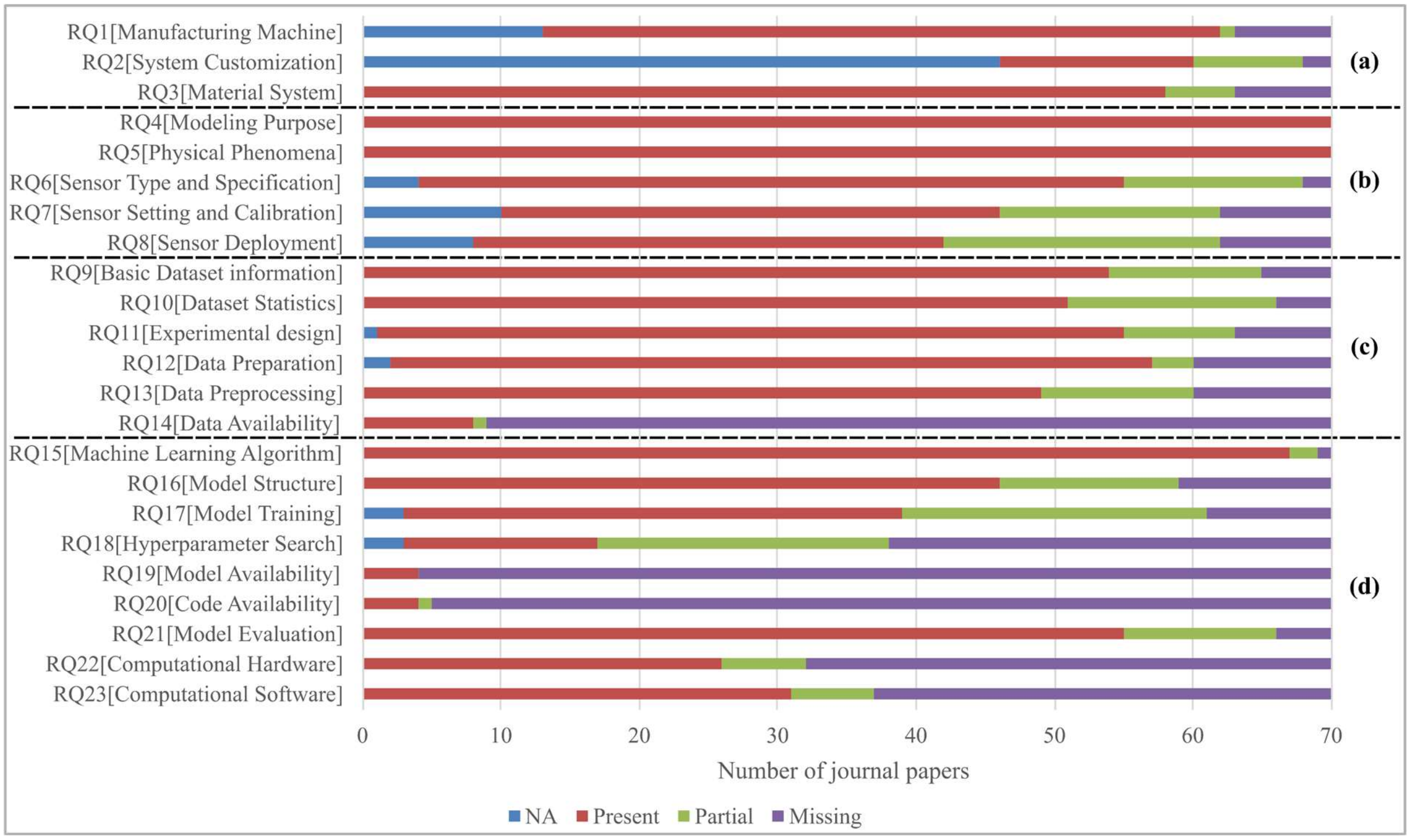

**Figure 12: Reproducibility survey results: (a) manufacturing system-related; (b) sensing system-related; (c) dataset-related; and (d) model-related.**

This survey revealed notable reproducibility problems in ML-based IIPQ-AM. Most authors fulfilled the manufacturing system-related reproducibility requirements owing to their professional AM background. The authors are aware of the sensing system-related and dataset-related reproducibility requirements but lack expertise in sensors, data analytics, and data handling. The partial and missing information on model-related reproducibility indicates a serious lack of ML understanding and expertise. There is an urgent need for reproducibility improvement in this domain, where the proposed reproducibility investigation pipeline and checklist could be helpful.

## 6. CONCLUSIONS

This paper proposed a reproducibility investigation pipeline and a reproducibility checklist for ML-based IIPQ-AM systems. Based on the CRISP methodology, the investigation pipeline guides readers through reproducing an original ML-based IIPQ-AM system. During the reproduction process, the proposed reproducibility checklist helps identify any missing reproducibility information. Additionally, the checklist assists original authors in ensuring that they provide all the necessary information for system reproduction in their publications. Two case studies were implemented to validate the proposed method, demonstrating that the reproduced models achieved performances comparable to those reported in the original manuscripts. The limitation of the proposed checklist (Version 1.1) and pipeline is that they are designed with a limited scope. This checklist does not cover several other ML applications in AM, including ML-assisted design and postprocessing. The checklist must therefore be expanded and adjusted to adapt to other ML applications in AM. This could be done by inviting experts in these fields to contribute their experience in enhancing the applicability of the checklist for the whole ML-assisted AM. The enhanced versions can be shared with relevant communities and venues for disseminating applied ML research in AM. Moreover, the SciHIT IE platform can be improved based on the modified versions of the checklist. In the future, more researchers can utilize the reproducibility investigation pipeline and checklist to enhance reproducibility in the AM domain. The proposed methods can be revised and adopted in a wide range of research domains where CPSs are involved.

## FUNDING DETAILS

Jiarui Xie received funding from Graduate Excellence Award (Grant# 00157) and McGill Engineering Doctoral Award (MEDA) fellowship of the Faculty of Engineering at McGill University. Jiarui Xie also received funding from Mitacs Accelerate Program (Grant# IT13369). Mutahar Safdar received funding from National Research Council Canada (Grant# NRC INT-015-1).

## ACKNOWLEDGMENTS

We would like to extend our sincere thanks to Dr. Kazem Fayazbakhsh and Aditya Saluja for generously sharing their data and providing invaluable information that was critical for completing the case study. Their contributions were essential to the success of this work.

## DISCLOSURE STATEMENT

The authors report there are no competing interests to declare.

## DATA AVAILABILITY STATEMENT

The data that support the findings of this study are available from the corresponding author, Prof. Yaoyao Fiona Zhao, upon reasonable request.

## AUTHOR CONTRIBUTIONS

**Jiarui Xie:** Conceptualization, Data curation, Methodology, Software, Validation, Formal analysis, Investigation, Writing - Original Draft, Visualization. **Mutahar Safdar:** Conceptualization, Data curation, Methodology, Software, Writing - Original Draft. **Andrei Mircea:** Conceptualization, Data curation, Software. **Bi Cheng Zhao:** Validation, Visualization. **Yan Lu:** Conceptualization, Methodology, Writing - Review & Editing, Supervision. **Hyunwoong Ko:** Conceptualization, Writing - Review & Editing, Supervision. **Zhuo Yang:** Conceptualization, Writing - Review & Editing, Resources. **Yaoyao Fiona Zhao:** Conceptualization, Resources, Writing - Review & Editing, Supervision, Funding acquisition.

**REPRODUCIBILITY CHECKLIST**

**Additive manufacturing process monitoring and quality prediction systems**

**Jiarui Xie, Mutahar Safdar, Andrei Mircea, Yan Lu, Hyunwoong Ko, Zhuo Yang, Yaoyao Fiona Zhao**

**Version 1.1 – clean version**

***Business Understanding***

**☐ RQ1[Manufacturing System]. Is the manufacturing system encompassing process technology and associated hardware component(s) reported?**

**☐ RQ2[System Customization]. If applicable, is customization of the manufacturing system reported?**

**☐ RQ3[Material System]. Are characteristics of the material system reported?**

**☐ RQ4[Modeling Purpose]. Is the machine learning modeling purpose discussed?**

**☐ RQ5[Physical Phenomena]. Are the physical phenomena being captured by the sensing system introduced?**

***Data Understanding***

**☐ RQ6[Sensor Type and Specification]. Are sensor type(s) and specification(s) provided?**

**☐ RQ7[Sensor Setting and Calibration]. Are the sensor settings or calibration details discussed?**

**☐ RQ8[Sensor Deployment]. Are the sensor deployment details discussed?**

**☐ RQ9[Basic Dataset information]. Are the data modalities, formats, and types introduced?**

**☐ RQ10[Dataset Statistics]. Are the relevant statistics provided?**

**☐ RQ11[Experimental design]. Are the experimental design settings for data acquisition introduced?**

***Data Handling***

**☐ RQ12[Data Preparation]. Are the data preparation techniques discussed, if any?**

**☐ RQ13[Data Preprocessing]. Are the data preprocessing techniques discussed, if any?**

**☐ RQ14[Data Availability]. Is there a link to a downloadable version of the dataset?**

***Modeling***

**☐ RQ15[Machine Learning Algorithm]. Is there a clear description of the machine learning algorithm?**

**☐ RQ16[Model Structure]. Are the details of the model described?**

**☐ RQ17[Model Training]. Are the details of model training described?**

**☐ RQ18[Hyperparameter Search]. If applicable, are hyperparameter search and selection procedures explained?**

**☐ RQ19[Model Availability]. Is the final trained model shared?**

**☐ RQ20[Code Availability]. Is the code open access or provided?**

***Evaluation***

**☐ RQ21[Model Evaluation]. Are the methods and metrics of model selection illustrated?**

***Deployment***

**☐ RQ22[Computational Hardware]. Is the computational hardware described?**

**☐ RQ23[Computational Software]. Is the computational software described?**